\documentclass[article,nofootinbib,10pt,floats,aps]{revtex4}

\usepackage{amsmath}
\usepackage{amssymb}
\usepackage{graphicx}
\usepackage{epsfig}
\usepackage{appendix}
\usepackage{bm}
\usepackage{units}
\usepackage{subfigure}
\usepackage{hyperref}

\newcommand{\be}{\begin{equation}}
\newcommand{\ee}{\end{equation}}
\newcommand{\bea}{\begin{eqnarray}}
\newcommand{\eea}{\end{eqnarray}}
\newcommand{\ben}{\begin{enumerate}}
\newcommand{\een}{\end{enumerate}}
\newcommand{\bit}{\begin{itemize}}
\newcommand{\eit}{\end{itemize}}

\newcommand{\la}[1]{\label{#1}}
\newcommand{\Eq}[1]{Eq.~(\ref{#1})}

\newcommand{\Sec}[1]{Sec.~\ref{#1}}

\newcommand{\Fig}[1]{Fig.~\ref{#1}}

\def\nn{\nonumber \\ }

\def\lsim{\mathrel{\rlap{\lower4pt\hbox{\hskip1pt$\sim$}} \raise1pt\hbox{$<$}}}
												
\def\gsim{\mathrel{\rlap{\lower4pt\hbox{\hskip1pt$\sim$}} \raise1pt\hbox{$>$}}}

\newcommand{\vv}[1]{\mathbf #1}						

\newcommand{\bert}{\raise-0.45mm\hbox{\Large$\Box$}}	

\begin{document}

\preprint{}

\title{The Sun's position in the sky}
 
\author{Alejandro Jenkins}\email{jenkins@hep.fsu.edu}

\affiliation{High Energy Physics, 505 Keen Building, Florida State University, Tallahassee, FL 32306-4350, USA}

\date{Aug. 2012, last revised Mar. 2013; published in Eur.\ J.\ Phys.\ {\bf 34}, 633 (2013)}


\begin{abstract}

We express the position of the Sun in the sky as a function of time and the observer's geographic coordinates.  Our method is based on applying rotation matrices to vectors describing points on the celestial sphere.  We also derive direct expressions, as functions of date of the year and geographic latitude, for the duration of daylight, the maximum and minimum altitudes of the Sun, and the cardinal directions to sunrise and sunset.  We discuss how to account for the eccentricity of the Earth's orbit, the precessions of the equinoxes and the perihelion, the size of the solar disk, and atmospheric refraction.  We illustrate these results by computing the dates of ``Manhattanhenge'' (when sunset aligns with the east-west streets on the main traffic grid for Manhattan, in New York City), by plotting the altitude of the Sun over representative cities as a function of time, and by showing plots (``analemmas'') for the position of the Sun in the sky at a given hour of the day. \\

{\it Keywords:} celestial sphere, rotation matrices, calendar, equation of the center, equation of time, precession, Manhattanhenge \\

{\it PACS:}
95.10.Km,		
02.40.Dr		

\end{abstract}

\maketitle

\tableofcontents

\section{Introduction}
\la{sec:intro}

This article will show how to compute the position of the Sun in the sky, for any given location on the surface of the Earth, at any given time.  Our method is based on describing the position of the Sun on the celestial sphere (a concept that should be very familiar to amateur astronomers) and performing several coordinate rotations on that sphere.  The idea is to begin with the {\it ecliptic} reference frame, in which the position of the Sun during the year is most easily and directly expressed, and to end with a {\it terrestrial} reference frame, defined with respect to an observer standing at a given point on the surface of the Earth, at a given time.  The mathematical training needed to understand this derivation is that which a student should have after a first course in linear algebra, since rotations will be described by matrices acting on three-dimensional vectors.  Familiarity with the transformation between rectangular (``Cartesian'') and spherical coordinates will be helpful, but shall not be assumed.

This work will allow us to arrive at mathematical expressions for the Sun's altitude above the horizon and for its geographic azimuth (i.e., its compass bearing), as functions of time and geographic location.  With an additional bit of geometry, we also obtain direct expressions, as functions of latitude and date, for the maximum and minimum solar altitudes, the number of continuous hours of daylight, and the cardinal directions of sunrise and sunset.  We will illustrate these formulas by plotting them for representative cities.

This pedagogical discussion also provides an opportunity to mention several interesting issues in celestial mechanics, such as Kepler's ``equation of the center,'' and the precessions of the equinoxes and the perihelion.  Another issue of astronomical interest that will be discussed is how the refraction of light, as it passes obliquely through the Earth's atmosphere, affects the apparent altitude of a celestial object.  We will also describe the phenomenon of ``Manhanttanhenge,'' when pedestrians in the borough of Manhattan, in New York City, may see the sunset in between the skyscrapers.  Finally, we illustrate the concepts of the ``equation of time'' and the analemma.

The purpose of this article is to give a self-contained, analytic characterization of the Sun's position in the sky, suitable for students without specialized training in astronomy or geodesy.  Astronomical and geodetic jargon will be avoided, or confined to footnotes, except insofar as it contributes to the argument's precision and clarity.  Angles will generally be expressed in radians and written as dimensionless numbers.  Geographic latitudes and longitudes, as well as solar altitudes and azimuths, will also be expressed in degrees (identified by a superscript $\circ$) when convenient.  In some cases, arc minutes (defined as sixtieths of a degree and identified by the symbol $'$) will also used.  In terms of notation, the guiding concern will be to achieve as much simplicity as possible without departing too far from the established usage.\footnote{For instance, some simplification could be achieved by working with the geographic co-latitude (i.e., the complement of the latitude), but I prefer to avoid this in deference to the widespread and long-established usage.}

The {\it Mathematica} notebook used to compute the solar altitudes and azimuths, and to produce the corresponding plots, is included with this arXiv submission as an ancillary file ({\tt SunPosition.nb}).  Interested readers are encouraged to use this notebook to explore the derivations in this article, extending or modifying the computations as they might see fit.

Computer codes are readily available on the Internet to find the position of the Sun in the sky (see, e.g., \cite{NOAA}).  The standard reference used in designing programs that compute the positions of celestial objects as functions of time (``ephemerides'') is \cite{Meeus}.  Almanacs such as \cite{almanac} also provide accurate values and formulas.  The Sun's position in the sky has recently been treated analytically in \cite{Probst,Sproul,Khavrus}, but these discussions make several approximations that we will avoid, and the overlap with the material covered here is only partial.

\section{Spherical coordinates for the Sun}
\la{sec:spherical}

\begin{figure} [t]
\begin{center}
	\includegraphics[width=0.37 \textwidth]{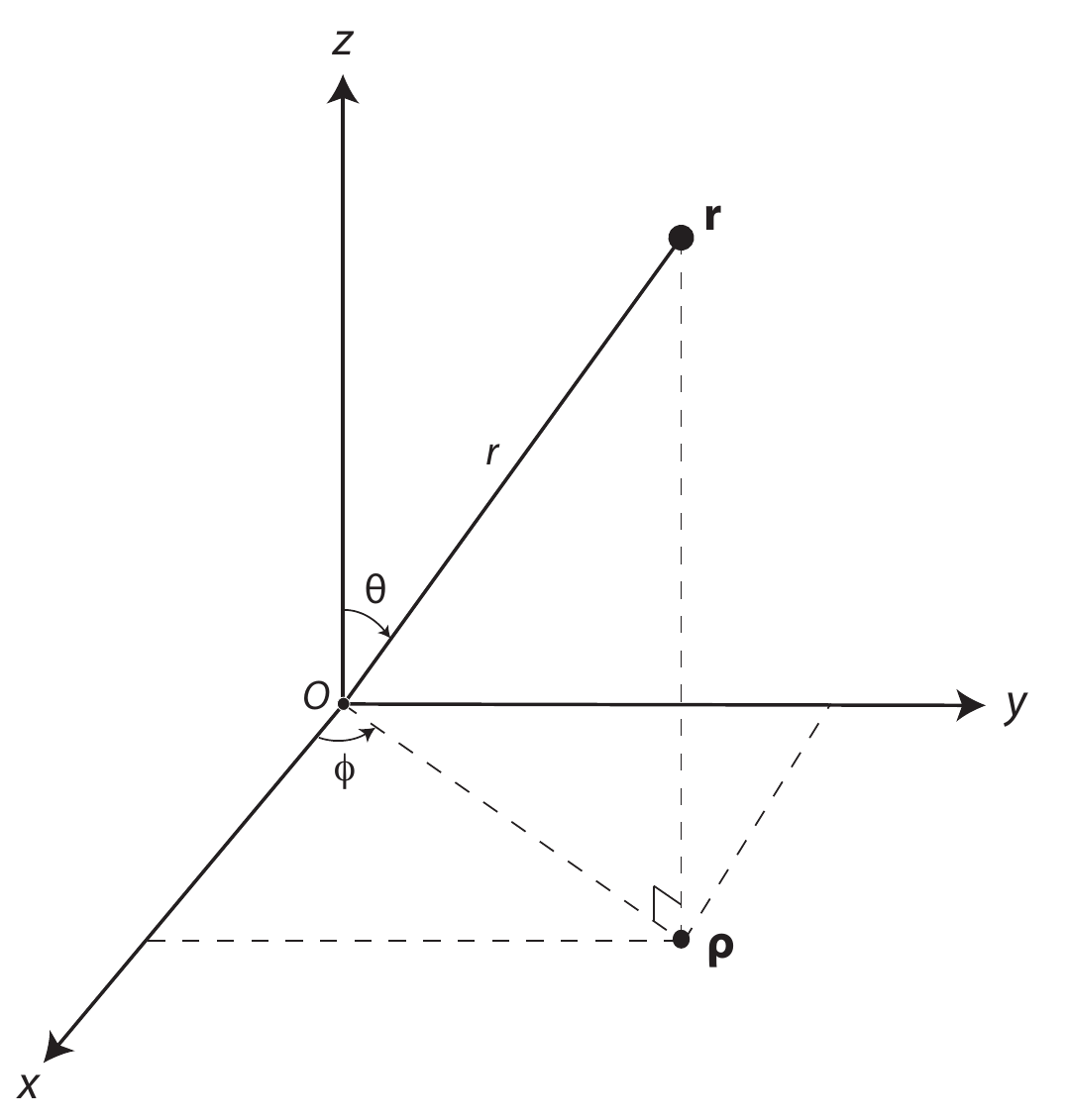}
\end{center}
\caption{\small In spherical coordinates, a three-dimensional vector $\vv r$ is expressed in terms of a radial distance $r$, a polar angle $\theta$, and an azimuthal angle $\phi$.  The vector {\boldmath $\rho$} is the projection of $\vv r$ onto the $x$-$y$ plane.  In terms of the rectangular coordinates, $x = r \sin \theta \cos \phi$, $y = r \sin \theta \sin \phi$, and $z = r \cos \theta$.\la{fig:spherical}}
\end{figure}

A point in three-dimensional space may be characterized by the spherical coordinates $(r, \theta, \phi)$, where $r$ is the radial distance, $\theta$ is the polar angle, and $\phi$ is the azimuthal angle.  In terms of the rectangular coordinates $(x, y, z)$, we have
\be
\vv r = \left( \begin{array}{c} x \\ y \\ z \end{array} \right) = \left( \begin{array}{c} r \sin \theta \cos \phi  \\ r \sin \theta \sin \phi \\ r \cos \theta \end{array} \right)~,
\la{eq:spherical}
\ee
as illustrated in \Fig{fig:spherical}.

The {\it celestial sphere} is an imaginary spherical surface, sharing a center with the Earth's globe, and with a very large, indefinite radius.  The positions of the stars, planets, and other heavenly bodies are characterized by their radial projection onto this surface.  The largeness of the radius of the celestial sphere, compared to the radius of the Earth, allows us, when convenient, to picture it as centered at the position of an observer standing on the Earth's surface, rather than at the center of the Earth.

For simplicity, we take the radius $r$ of the celestial sphere to be equal to 1 (in undetermined units).  We shall use a subscript $\odot$ (the astronomical symbol for the Sun) to indicate that a vector or a coordinate thereof refers to the position of the Sun.

\subsection{Ecliptic frame}
\la{sec:ecliptic}

From the Earth, the Sun appears to move, against the background of the distant stars, along a great circle on the celestial sphere called the {\it ecliptic}.\footnote{The ecliptic is sometimes defined as the plane of the Earth's orbit around the Sun.  The circle that we call the ``ecliptic'' is the intersection of that plane with the celestial sphere.}  We will therefore start by working in an ``ecliptic frame,'' in which the position of the distant stars is fixed,\footnote{For this reason the distant stars, which form the constellations, are also referred to as the ``fixed stars.''} and in which the polar angle of the Sun is always $\theta_\odot = \pi / 2$, whereas the azimuthal angle $\phi_\odot$ varies over the course of the year, as shown in \Fig{fig:ecliptic}.  If the Earth's orbit were perfectly circular, then $\phi_\odot$ would increase at a constant rate, completing a full revolution in a year.  In \Sec{sec:center} we will see how to account for the fact that the Earth's orbit is slightly elliptical, but for now we will simply express the azimuthal angle of the Sun as a function of the time $t$.  We therefore express the position of the Sun, in the ecliptic frame of reference, as:
\be
\vv r_\odot (t) = \left( \begin{array}{c} \cos \phi_\odot (t) \\ \sin \phi_\odot (t) \\ 0 \end{array} \right)~.
\la{eq:ecliptic}
\ee

\begin{figure} [t]
\begin{center}
	\includegraphics[width=0.47 \textwidth]{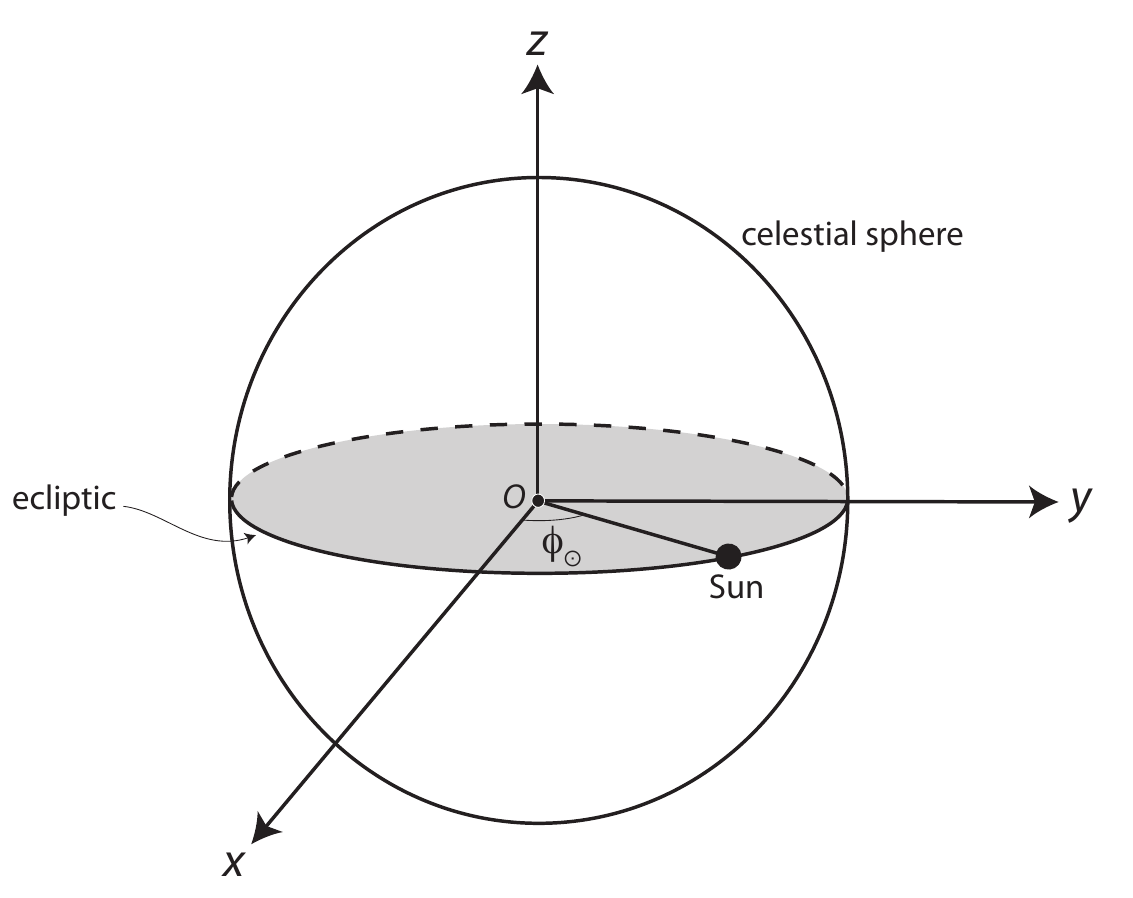}
\end{center}
\caption{\small The Sun moves along the ecliptic during the course of the year.  In the ecliptic frame of reference, the Sun's polar angle is fixed, $\theta_\odot = \pi / 2$, while the azimuthal angle $\phi_\odot$ increases with time at an approximately constant rate of $2 \pi$ per year.\la{fig:ecliptic}}
\end{figure}

\subsection{Equatorial frame}
\la{sec:equatorial}

The axis of rotation of the Earth is tilted with respect to the ecliptic frame by an angle of obliquity
\be
\varepsilon = 23.44^\circ = 0.4091 ~.
\la{eq:obliquity}
\ee
It is therefore convenient to change coordinates to an ``equatorial frame,'' by rotating about the $x$-axis by an angle $\varepsilon$, as shown in \Fig{fig:equatorial}(a), so that the new $z'$-axis coincides with the Earth's axis of rotation.  The motion of the Sun in this equatorial frame is illustrated in \Fig{fig:equatorial}(b), in which the celestial north pole is labelled $P$ and the celestial south pole $\overline P$.  The ecliptic intersects the celestial equator at two points, $e$ and $\bar e$, known as the {\it equinoxes}.  At $e$ the Sun crosses the equator from south to north, and this is therefore known as the northward equinox (or ``first equinox'', since it occurs first in the calendar year).  Conversely, $\bar e$ is known as the southward, or second equinox.  The points of maximum displacement between the position of the Sun and the celestial equator are known as {\it solstices}, and are marked in \Fig{fig:equatorial}(b) by $s$ and $\bar s$.

The position of Sun in the equatorial coordinate frame is given by:
\bea
\vv r'_\odot =
\left( \begin{array}{c} x'_\odot \\ y'_\odot \\ z'_\odot \end{array} \right) &=&
\left( \begin{array}{c} \sin \theta'_\odot \cos \phi'_\odot \\ \sin \theta'_\odot \sin \phi'_\odot \\ \cos \theta'_\odot \end{array} \right) \nn
&=&
\left( \begin{array}{r r r}
1 & 0 & 0 \\
0 & \cos \varepsilon & - \sin \varepsilon \\
0 & \sin \varepsilon & \cos \varepsilon
\end{array} \right)
\left( \begin{array}{c} \cos \phi_\odot  \\ \sin \phi_\odot \\ 0 \end{array} \right) = 
\left( \begin{array}{c} \cos \phi_\odot  \\ \cos \varepsilon \sin \phi_\odot \\ \sin \varepsilon \sin \phi_\odot \end{array} \right) ~.
\la{eq:equatorial}
\eea
The polar angle for the Sun in this equatorial reference frame is therefore
\be
\theta'_\odot = \arccos z'_\odot = \arccos \left( \sin \varepsilon \sin \phi_\odot \right)~.
\la{eq:polar-sun}
\ee

In the astronomical literature, the value of $\phi_\odot$, measured with respect to the first equinox $e$, is called the Sun's ``ecliptic longitude.''  The corresponding $\phi'_\odot$ in the equatorial frame is called the Sun's ``right ascension,'' while $\pi / 2 - \theta'_\odot$ is its ``declination.''\footnote{It is common to approximate the Sun's declination as \hbox{$\pi / 2 - \theta'_\odot \simeq \varepsilon \sin \phi_\odot$}; see, e.g., \cite{Khavrus}.  This is accurate to within $0.26^\circ$ for the value of $\varepsilon$ in \Eq{eq:obliquity}, but rather obscures the geometry involved, as Sproul comments in \cite{Sproul}.}  For a full discussion of the celestial sphere and of the coordinate systems that astronomers use to characterize points on it, see \cite{celestial-sphere}.

\begin{figure} [t]
\begin{center}
	\subfigure[]{\includegraphics[width=0.3 \textwidth]{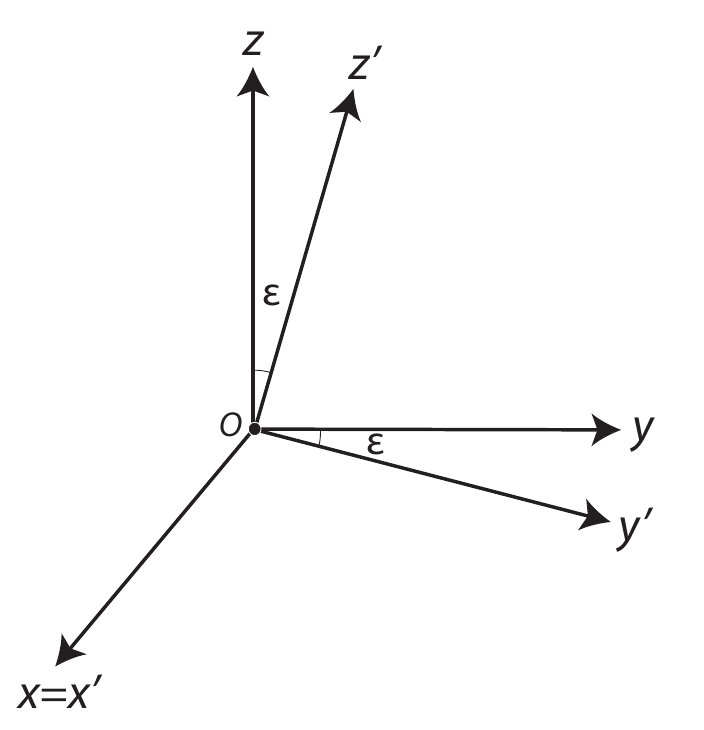}} \hskip 2.5 cm
	\subfigure[]{\includegraphics[width=0.42 \textwidth]{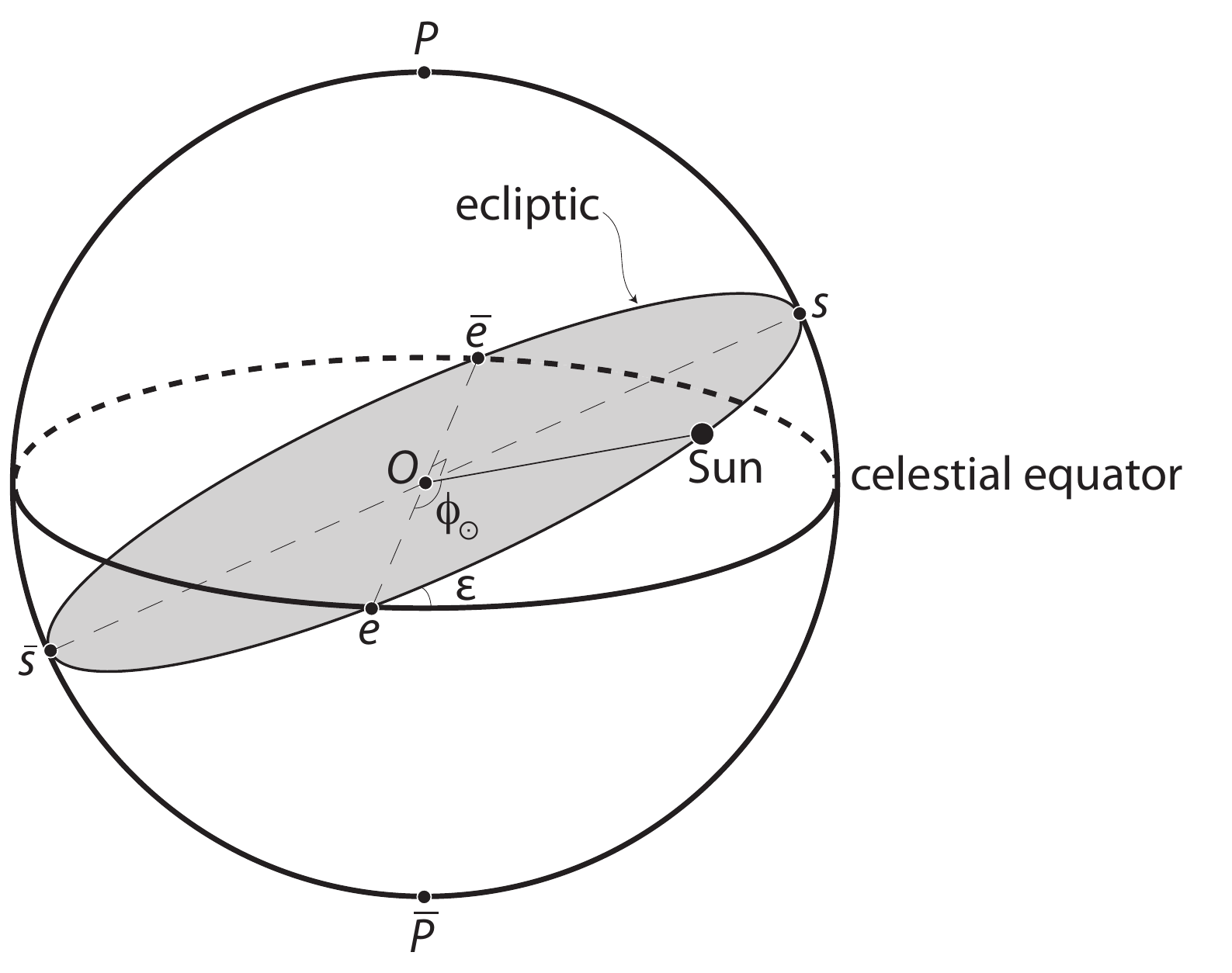}}
\end{center}
\caption{\small (a): We may transform from the ecliptic to the equatorial reference frame by rotating along the $x$-axis by an angle equal to the obliquity $\varepsilon$, given in \Eq{eq:obliquity}, so that the new axis $z'$-axis is also the axis of the Earth's rotation.  (b):  The motion of the Sun along the ecliptic, as seen in the new equatorial reference frame.  Point $P$ marks the celestial north pole and $\overline P$ the celestial south pole.  The northward and southward equinoxes are marked by $e$ and $\bar e$, respectively.  The northern and southern solstices are indicated by $s$ and $\bar s$, respectively.\la{fig:equatorial}}
\end{figure}

\subsection{Terrestrial frame}
\la{sec:terrestrial}

Seen from the Earth, objects in the sky rotate azimuthally in the equatorial frame (i.e., about the $z'$-axis), with constant angular velocity
\be
\omega = \frac{2\pi}{23.9345 ~\hbox{hours}}~,
\la{eq:omega}
\ee
where $23.9345$ hours is the duration of the ``sidereal day,'' equal to the amount of time that it takes the Earth to complete one rotation about its axis (and therefore also for a distant star to return to the same position in the sky).  This is slightly less than the ``mean solar day'' of 24 hours, because of the Sun's motion along the ecliptic during the course of one sidereal day.

To characterize the position of the Sun, as seen from a point on the surface of the Earth, we must also adjust for the geographic latitude $L$.  We can achieve this by rotating about the $x$-axis by an angle equal to the co-latitude $\pi / 2 - L$.  The transformation from the equatorial frame to the terrestrial frame therefore gives: 
\bea
\vv r''_\odot =
\left( \begin{array}{c} x''_\odot \\ y''_\odot \\ z''_\odot \end{array} \right) &=&
\left( \begin{array}{c} \sin \theta''_\odot \cos \phi''_\odot \\ \sin \theta''_\odot \sin \phi''_\odot \\ \cos \theta''_\odot \end{array} \right) \nn
&=& \left( \begin{array}{r r r}
1 & 0 & 0 \\
0 & \sin L & - \cos L \\
0 & \cos L & \sin L
\end{array} \right)
\left( \begin{array}{r r r}
\cos \left[ \omega (t - t_0) \right] & \sin \left[ \omega (t - t_0) \right] & 0 \\
-\sin \left[ \omega (t - t_0) \right] & \cos \left[ \omega (t - t_0) \right] & 0 \\
0 & 0 & 1
\end{array} \right)
\left( \begin{array}{c} \cos \phi_\odot  \\ \cos \varepsilon \sin \phi_\odot \\ \sin \varepsilon \sin \phi_\odot \end{array} \right)~,
\la{eq:terrestrial}
\eea
where $t - t_0$ is the interval during which the Earth has rotated, measured with respect to a reference time $t_0$.\footnote{Note that the signs of the off-diagonal $\pm \sin[\omega(t-t_0)]$ entries in the corresponding rotation matrix in \Eq{eq:terrestrial} reflect the fact that the Earth's rotation displaces the Sun in an azimuthal direction {\it opposite} to that of the Sun's yearly motion along the ecliptic.  This is the reason why the mean solar day of 24 hours is {\it longer} than the sidereal day of 23.9345 hours: the extra 4 minutes of rotation are need to compensate for the change in $\phi_\odot$ in \Eq{eq:ecliptic}.}  We will discuss how to choose the value of $t_0$ (which will depend on the geographic longitude $\ell$) in \Sec{sec:longitude}.

By \Eq{eq:terrestrial}, the altitude (or ``elevation'') of the Sun above the horizon, as a function of the latitude and the time $t$, is
\bea
\alpha_\odot (L, t) = \frac{\pi}{2} - \theta''_\odot (L, t) &=& \arcsin [z''_\odot (L, t)] \nn
&=& \arcsin \left( - \cos L \cdot \cos [\phi_\odot (t)] \cdot \sin \left[ \omega (t - t_0) \right]
+ \cos L \cdot \cos \varepsilon \cdot \sin [\phi_\odot (t)] \cdot \cos \left[ \omega (t - t_0) \right] \right. \nn
&&  \hskip 1.5 cm \left. +  \sin L \cdot \sin \varepsilon \cdot \sin [\phi_\odot (t)] \right)~.
\la{eq:altitude}
\eea
When $\alpha_\odot = 0$, the Sun is either rising or setting.  When $\alpha_\odot = \pi / 2$, the Sun is directly overhead, at the ``zenith'' (this can occur only at tropical latitudes $-\varepsilon \leq L \leq \varepsilon$).  

Meanwhile, the azimuthal angle $\phi''_\odot$ can be computed from \Eq{eq:terrestrial}, using the relation
\be
\tan \phi''_\odot = \frac{y''_\odot}{x''_\odot}~.
\la{eq:terr-azimuth}
\ee
In \Sec{sec:longitude} we will work out the relation between this $\phi''_\odot$ and the cardinal directions (North, East, South, and West).

\section{Astronomical adjustments}
\la{sec:astronomical}

For some purposes, it may be acceptable to approximate the angle $\phi_\odot (t)$ as increasing linearly and completing a full revolution in one year (as do the authors of \cite{Khavrus}). A more precise expression can be obtained from Kepler's first and second laws of planetary motion, which state that the Earth moves along an ellipse, with the Sun at a focus, while the line segment from the Sun to the Earth sweeps out equal areas in equal times.

\subsection{Equation of the center}
\la{sec:center}

The angle subtended by the line from the Sun to the Earth, with respect to the major axis of the elliptical orbit, is known to astronomers as the ``true anomaly'' and is usually represented by the letter $v$.  Finding $v$ as a function of time has no exact analytic solution,\footnote{Newton offered a rigorous proof that no analytic solution could exist, using concepts now associated with topology, long before topology was invented.  This fascinating proof (the first impossibility proof since the ancient Greeks) is discussed in \cite{Arnold,Chandra}.} but an expansion can be obtained, which converges rapidly for small orbital eccentricity $e$, known as the ``equation of the center:''
\be
v = M + 2 e \sin M + \frac{5}{4} e^2 \sin 2M + \frac{1}{12} e^3 \left( 13 \sin 3M - 3 \sin M \right) + {\cal O}(e^4)~.
\la{eq:center}
\ee
The ``mean anomaly'' in \Eq{eq:center} can be expressed as
\be
M = M_0 + M_1 t~,
\la{eq:mean}
\ee
with constant $M_{0,1}$; it would be equal to the angle $v$ for a perfectly circular orbit ($e=0$) of equal area to the true elliptical orbit; see \cite{Danby}.  The values of $v$ and $M$ in \Eq{eq:center} are measured with respect to the {\it perihelion}, which is the point of closest approach between the Earth and the Sun, as shown in \Fig{fig:orbit}.  The value of $2 \pi / M_1$ is slightly greater than one calendar year because of the slow precessions of the equinoxes and the perihelion, which we will discuss in \Sec{sec:precession}. 

\begin{figure} [t]
\begin{center}
	\includegraphics[width=0.6 \textwidth]{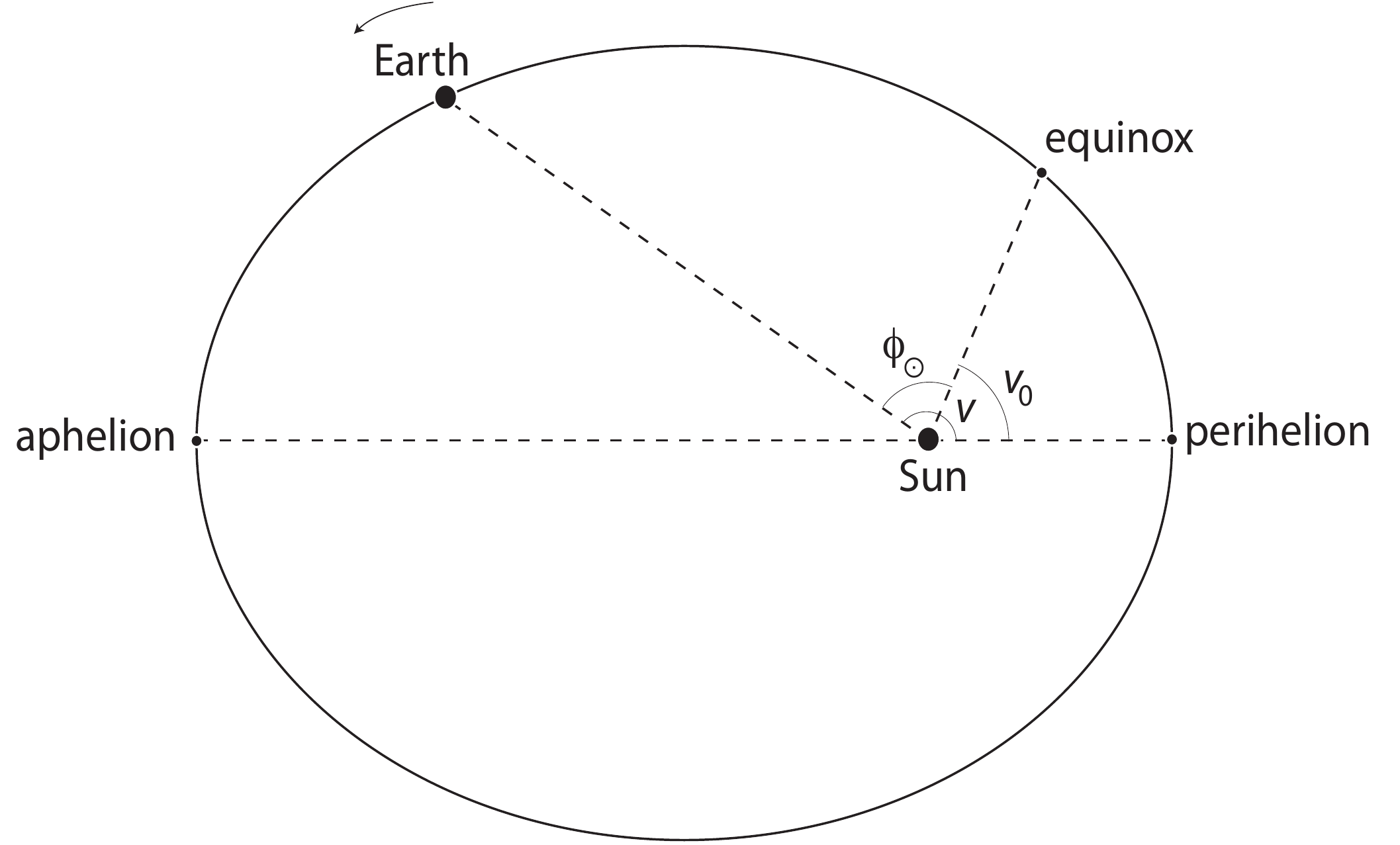}
\end{center}
\caption{\small Diagram of the Earth's orbit around the Sun.  The eccentricity is exaggerated for clarity.  The ``true anomaly'' $v$ is measured from perihelion, which is the point of closest approach between the Sun and the Earth.  The point of greatest separation between Sun and Earth is called the ``aphelion.''  We measure the ecliptic azimuthal angle of the Sun, $\phi_\odot$, from the Earth's position at the time of the first equinox.  Therefore $\phi_\odot = v - v_0$, where $v_0$ is the angular displacement between the perihelion and the equinox.\la{fig:orbit}}
\end{figure}

For our purposes it will be convenient to measure the angle $\phi_\odot$ from the first equinox of the year.  Therefore we let
\be
\phi_\odot = v - v_0 ~,
\la{eq:v0}
\ee
where $v_0$ is the angular displacement between the perihelion and the first equinox, as shown in \Fig{fig:orbit}.  Using the current astronomical data for the parameters $M_0$, $M_1$, $e$, and $v_0$ \cite{orbit-parameters}, we can write the equation of the center for the Earth as:
\be
M(t) = -0.0410 + 0.017202 \, t
\la{eq:mean-numbers}
\ee
and
\be
\phi_\odot (t) = - 1.3411 + M(t) + 0.0334 \sin [M(t)] + 0.0003 \sin [2M(t)]~,
\la{eq:center-numbers} 
\ee
where $t = 0$ corresponds to 1 January 2013, 0:00, Universal Coordinated Time (UTC), and $t$ is measured in mean solar days of 24 hours.

We may see from \Eq{eq:center-numbers} that the correction to $\phi_\odot$ introduced by the eccentricity of the Earth's orbit is small: less than $2^\circ$ at any given time of the year.  For a planet like Mercury, whose orbit is more eccentric and whose rotation is slower than the Earth's, the motion of the Sun in the sky is qualitatively different, as discussed in \cite{Mercury}.

\subsection{Precession of equinoxes and perihelion}
\la{sec:precession}

In the second century BCE, the Greek astronomer Hipparchos of Nicaea found that the positions of the equinoxes moved along the ecliptic (i.e., with respect to the distant stars) by about $1^\circ$ per century (the modern estimate is $1.38^\circ$ per century).  Newton correctly explained this as due to the tidal forces that the Moon and the Sun exert on the Earth, which is not perfectly spherical.  If the Earth did not spin, those tidal forces would pull the Earth's equatorial bulge onto the orbital plane of the corresponding perturbing body (i.e., of either the Moon or the Sun).  The Earth's spinning turns the action of that tidal torque into a {\it precession}, so that the axis of the Earth's rotation describes a cone, and the position of the celestial north pole therefore moves slowly along a circle, with respect to the constellations.\footnote{This slow change of the positions of the poles, equinoxes, and solstices, relative to the distant stars, implies that the signs of the Zodiac are not fixed with respect to the solar calendar.  For example, the ``Tropic of Cancer'' was so named because the position of the Sun at the time of the northern solstice used to lie within the constellation of Cancer, but today the northern solstice actually lies in Taurus.  The first equinox, which used to lie in Aries when the ancient Babylonians developed the calendar, has since shifted to Pisces and will move into Aquarius around the year 2,600.  This last circumstance has been the source of much mystical twaddle about the ``dawning of the Age of Aquarius.''}

The period of the precession of the Earth's axis is about 26,000 years.  Since the recurrence of the seasons depends on the periodicity of the equinoxes, rather than on the actual time it takes the Earth to go once around the Sun, the modern calendar is based on the ``mean tropical year,'' which is shorter than the sidereal year by about $20$ minutes (i.e., 1/26,000 of a sidereal year).

The position of the perihelion with respect to the distant stars also varies, but more slowly, with a period of about 112,000 years, which is equivalent to a displacement of about $0.32^\circ$ per century.  This precession results from perturbations to the motion of the Earth around the Sun caused by the gravitational pull of the Moon and the other planets, and to a lesser extent also by relativistic corrections to Newtonian gravity.\footnote{One of the most convincing early demonstrations of the validity of Einstein's theory of general relativity was that it explained the anomalous precession of the perihelion for the orbit of Mercury, which astronomers had until then failed to account for by the gravitational influence of the known planets; see \cite{dInverno} for a detailed discussion.}

The respective precessions of the equinox and the perihelion proceed in opposite directions along the ecliptic, causing the value of $v_0$ in \Eq{eq:v0} to {\it decrease} by about $1.7^\circ$ per century.\footnote{The quantity $2\pi - v_0$ is known to astronomers the ``longitude of perihelion.''}  Though for our purposes such precision is hardly justified, if we wished to take into account those precessions, we could make $v_0$ in \Eq{eq:v0} a time-dependent parameter.

\section{Duration of daylight}
\la{sec:daylight}

The computation only up to \Eq{eq:polar-sun} suffices to obtain a good estimate of the number of hours of daylight for a given day of the year, if we do not care for the precise time of sunrise and sunset.  Here the main approximation is that that the azimuthal angle of the Sun in the ecliptic frame, $\phi_\odot$, will be taken to be fixed during a given calendar date $d$.  For definiteness, let us say that $\phi_\odot$ is computed at noon for the date and location of interest, the corresponding time being translated to Universal Coordinated Time (UTC), for use in Eqs.~(\ref{eq:mean-numbers}) and (\ref{eq:center-numbers}).

\begin{figure} [t]
\begin{center}
	\subfigure[]{\includegraphics[width=0.5 \textwidth]{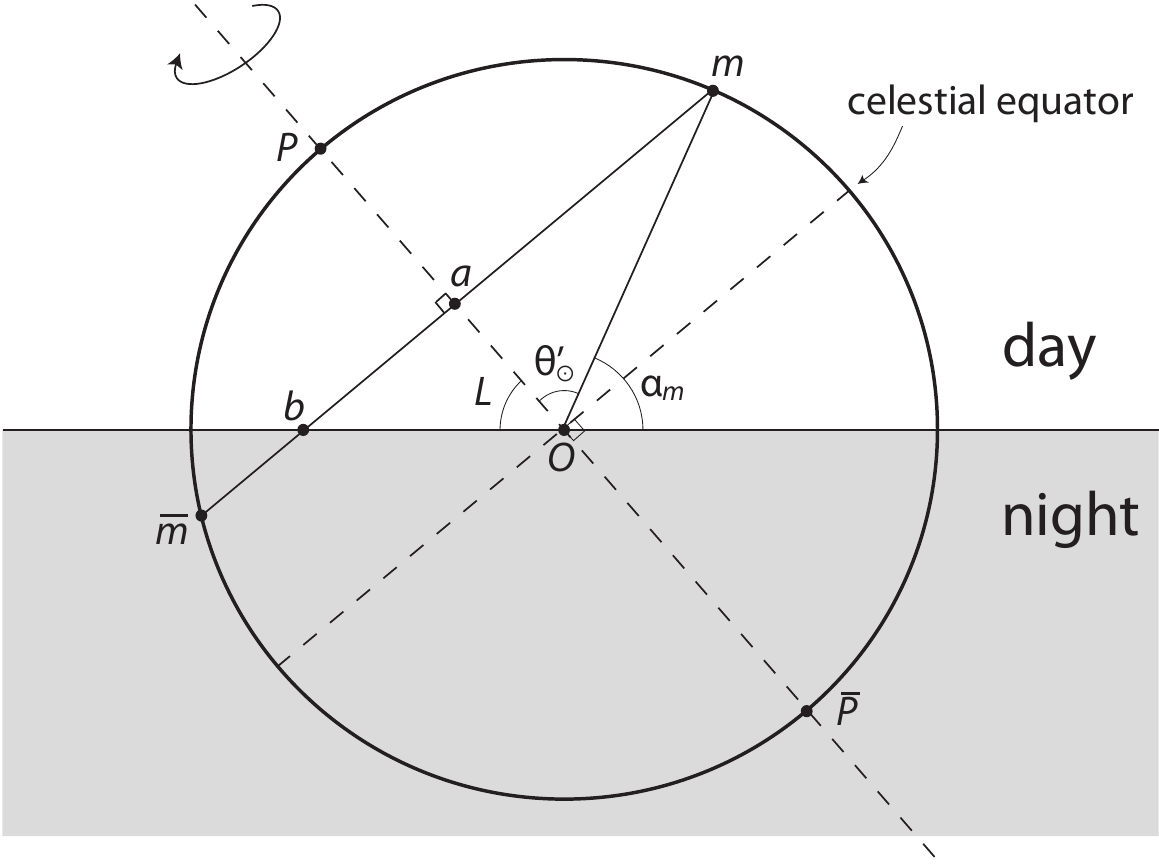}} \hskip 1.5 cm
	\subfigure[]{\includegraphics[width=0.325 \textwidth]{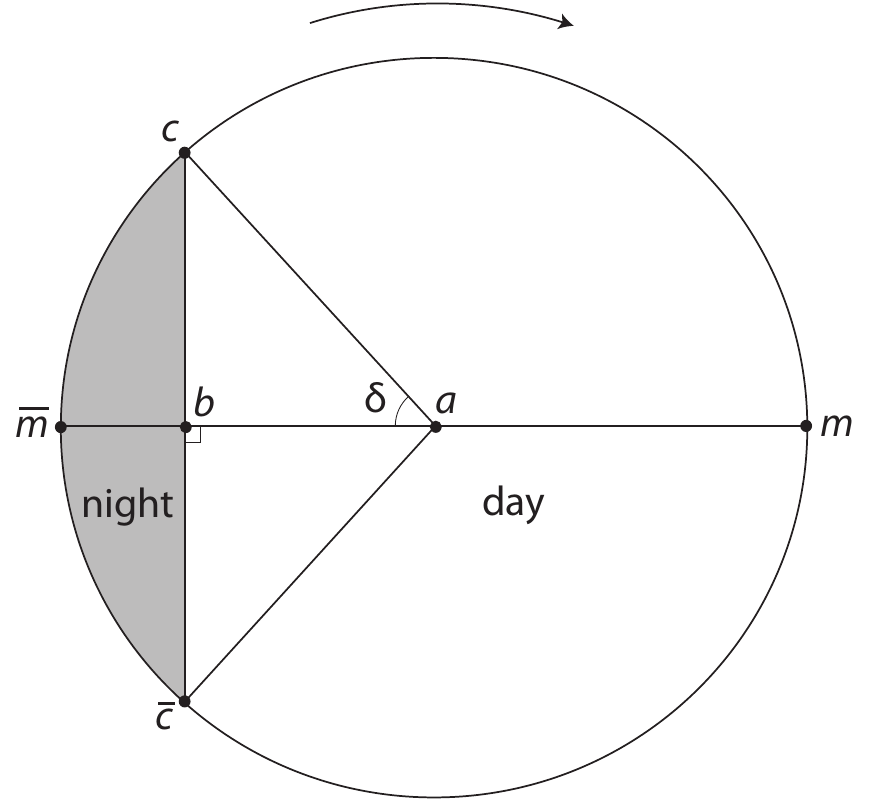}}
\end{center}
\caption{\small (a): Cross-section of the celestial sphere along the Earth's axis of rotation $P \overline P$, centered at the position $O$ of an observer at geographic latitude $L$.  The point $m$ corresponds to the maximum altitude of the Sun, and $\overline m$ to the minimum altitude.  (b): Cross-section of the celestial sphere, centered at point $a$ and perpendicular to the Earth's axis of rotation.  The point $c$ corresponds to sunrise and $\bar c$ to sunset.  The arrows show the direction in which the celestial sphere rotates with respect to the observer at $O$.\la{fig:daylight}}
\end{figure}

Figure \ref{fig:daylight}(a) shows a cross-section of the celestial sphere, parallel to the Earth's axis of rotation $P \overline P$.  As the sphere rotates about the observer at point $O$, the celestial pole $P$ maintains a fixed altitude, equal to the observer's geographic latitude $L$.\footnote{If we take $L$ to be positive for points on the northern hemisphere of the Earth, then $P$ is the north celestial pole, and $\overline P$ is the south celestial pole.  The opposite convention would be more convenient for observers in the southern hemisphere.}  Point $m$ marks the maximum altitude of the Sun, while point $\overline m$ marks its minimum altitude.

The path of the Sun in the sky corresponds to the circle $am$, shown in \Fig{fig:daylight}(b) (again, as long as we neglect the change in $\phi_\odot$, and therefore also in $\theta'_\odot$, during the course of one day).  This circle is a cross-section of the celestial sphere, perpendicular to the axis $P \overline P$ and parallel to the line $m \overline m$.

In terms of the angle $\delta$ in \Fig{fig:daylight}(b),\footnote{Astronomers call $\delta$ the Sun's ``local hour angle'' at the times of rising and setting.  See, e.g., \cite{Meeus-rising}.} the number of hours of daylight is simply
\be
H = 24 \left( 1 - \frac{\delta}{\pi} \right)~,
\ee
since the Sun moves uniformly along the circle $am$, with a period of 24 hours.\footnote{By making the period of rotation of the Sun about the celestial poles in \Fig{fig:daylight} equal to the mean solar day of 24 hours, rather than the sidereal day of 23.9345 hours, we are taking into account the average change in $\phi_\odot$ during the course of one day.}  Examining Figs.~\ref{fig:daylight}(a) and (b), we see that
\be
\delta = \arccos \frac{ab}{am} = \arccos \left( \tan L \cot \theta'_\odot \right)~.
\la{eq:delta}
\ee
Therefore we can express the number of hours of daylight as a function of geographic latitude and day of the year in the form:
\bea
H(L,d) &=& 24 \left[ 1 - \frac{\arccos \left( \tan L \cot \theta'_\odot (d) \right)}{\pi} \right] \nn
&=& 24 \left[ 1 - \frac{1}{\pi} \arccos \left( \tan L \frac{\sin \varepsilon \sin [\phi_\odot (d)]}
{\sqrt{1 - \sin^2 \varepsilon \sin^2 [\phi_\odot (d)]}} \right) \right]
\la{eq:hours}
\eea
(which agrees with the expression obtained in \cite{Khavrus}).

\begin{figure} [t]
\begin{center}
	\includegraphics[width=0.5 \textwidth]{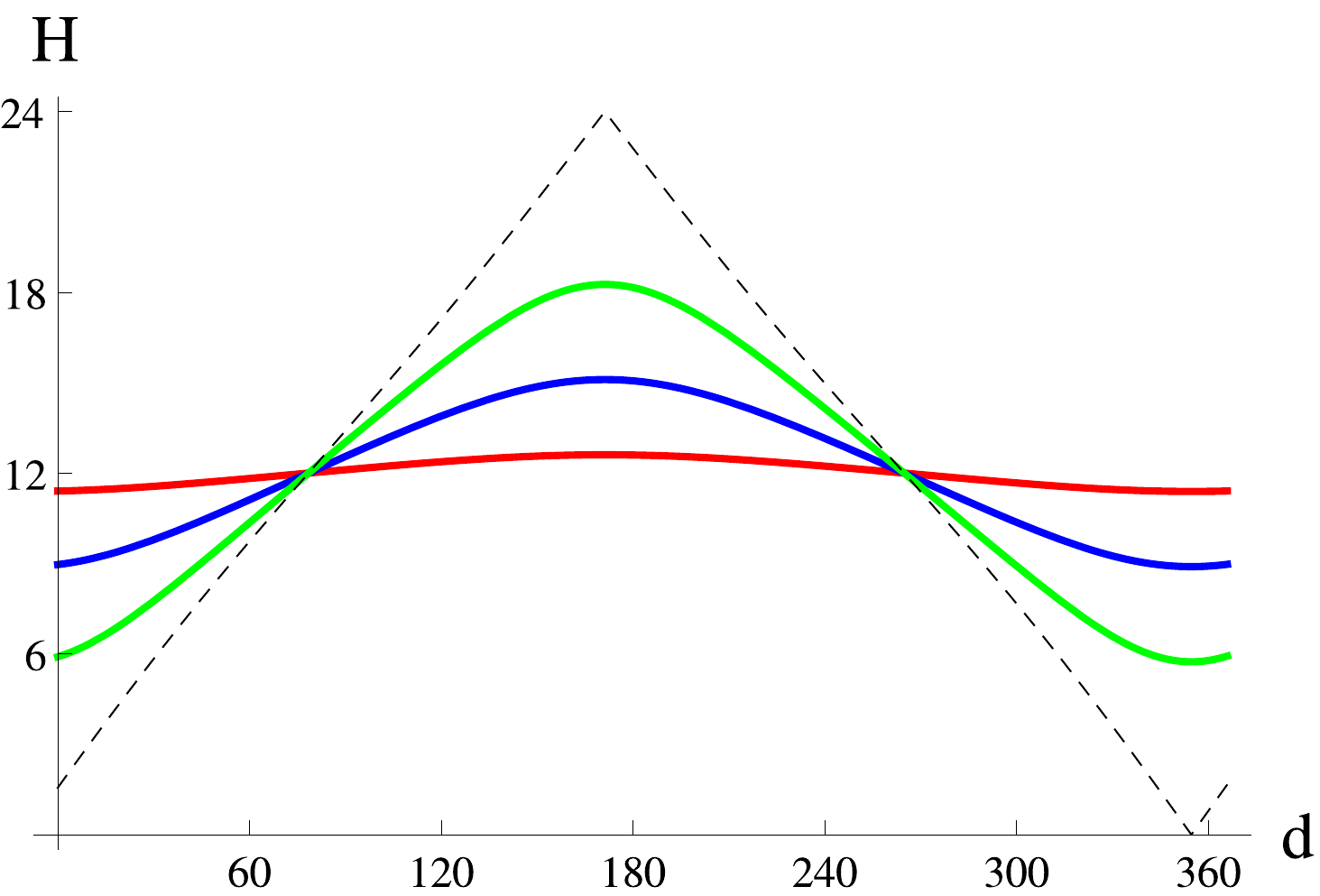}
\end{center}
\caption{\small Number of hours of continuous daylight $H$, as a function of the day of the year $d$ (starting on 1 January), computed using \Eq{eq:hours}, for: the latitude of Cartagena de Indias, Colombia, $10^\circ 24'$ N (red curve); the latitude of Boston, Massachusetts, USA, $42^\circ 21'$ N (blue curve); the latitude of Stockholm, Sweden, $59^\circ 20'$ N (green curve); and the Arctic Circle, $66^\circ 34'$ N (dashed black curve).\la{fig:daylight-hours}}
\end{figure}

Figure \ref{fig:daylight-hours} shows plots of $H$ as a function of the day of the year $d$, at the latitudes of Cartagena de Indias (Colombia), Boston (USA), Stockholm (Sweden), and the Arctic Circle, all in the northern hemisphere.  Note that, for the northern hemisphere, the midyear solstice (which occurs around 21 June, or $d=171$) is always the longest day, whereas it is the shortest day everywhere in the southern hemisphere.  Conversely, the year-end solstice (around 21 December, or $d = 354$) is always the longest day in the southern hemisphere and the shortest in the northern hemisphere.

\subsection{Maximum and minimum solar altitudes}
\la{sec:elevation}

In \Fig{fig:daylight}, it is easy to see that the maximum and minimum altitudes of the Sun on a given date, which we respectively label $\alpha_m$ and $\alpha_{\overline m}$, are:
\bea
\alpha_m (L, d)&=& \arcsin \left( \sin \left[ L + \theta'_\odot (d) \right] \right) \nn
\alpha_{\overline m} (L, d) &=& \arcsin \left( \sin \left[ L - \theta'_\odot (d) \right] \right) ~,
\la{eq:a-maxmin}
\eea
where $\theta'_\odot$ is given by \Eq{eq:polar-sun}.  These are the values between which the solar altitude of \Eq{eq:altitude} varies during the day.  The use of the arcsine function in \Eq{eq:a-maxmin} reflects the fact that an altitude must, by definition, be between $\pm \pi / 2$.

For tropical latitudes $-\varepsilon \leq L \leq \varepsilon$ it is {\it not} true that the Sun reaches its maximum altitude on the day of the solstice.  This is shown graphically in \Fig{fig:zenith}, which plots the maximum and minimum solar altitudes, given by \Eq{eq:a-maxmin}, as functions of the day of the year, for the latitudes of San Jos\'e, Costa Rica ($9^\circ 56'$ N) and Casablanca, Morocco ($33^\circ 32'$ N).  While the noonday Sun over Casablanca does reach its maximum altitude (\hbox{$\alpha_m = pi / 2 - L + \varepsilon$}) on the day of the midyear solstice, the Sun over tropical San Jos\'e reaches zenith ($\alpha_m = \pi / 2$) on two days, one before and the other after the solstice.  We shall compute these dates in \Sec{sec:sanjose}.

\begin{figure} [t]
\begin{center}
	\includegraphics[width=0.52 \textwidth]{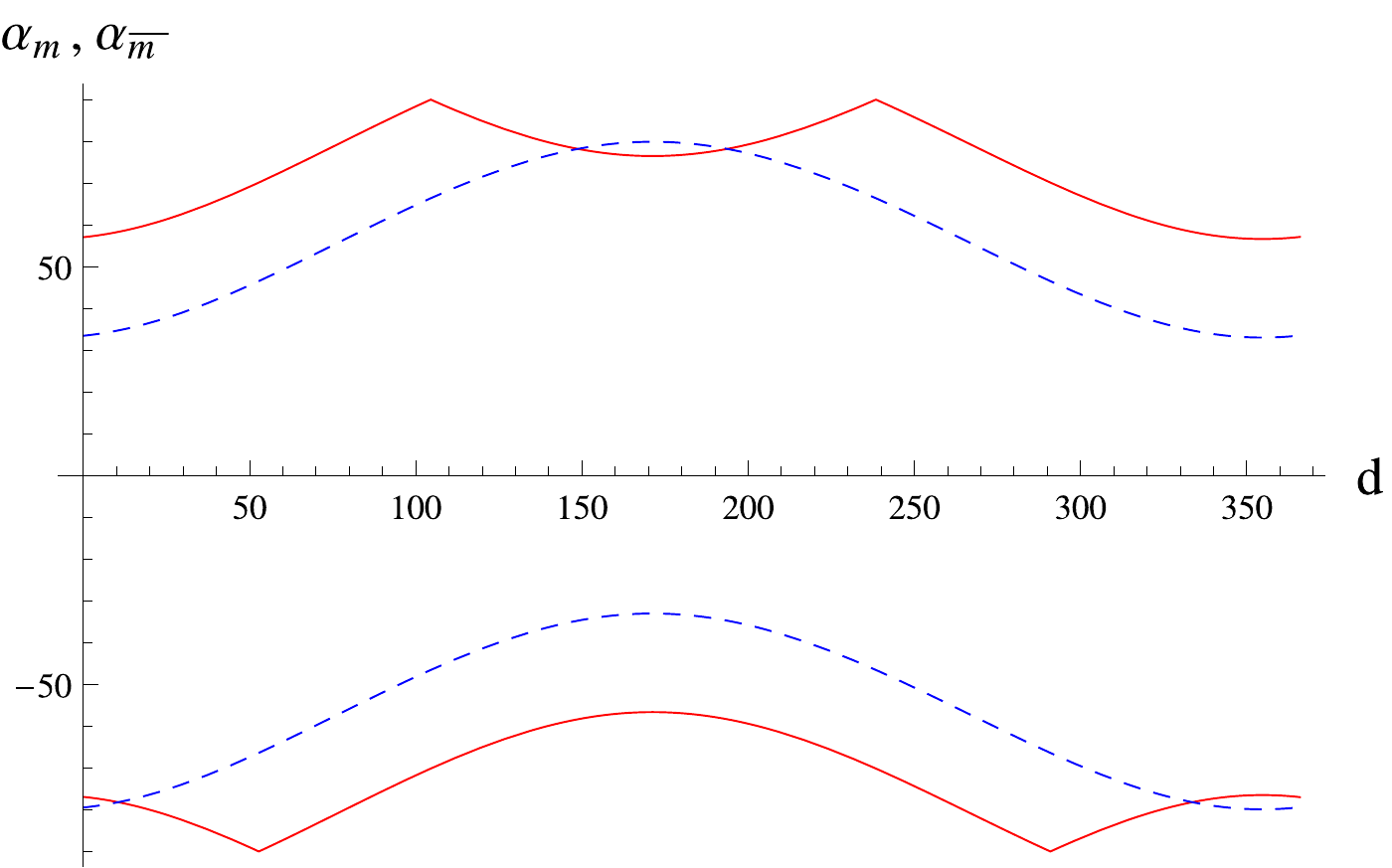}
\end{center}
\caption{\small The upper red curve corresponds to the maximum altitude of the Sun, $\alpha_m$, in degrees, as a function of the day of the year $d$ (starting on 1 January), at the latitude of San Jos\'e, Costa Rica ($9^\circ 56'$  N).  The lower red curve gives the minimum altitude $\alpha_{\overline m}$ at that same latitude.  The dashed blue curves give $\alpha_m$ and $\alpha_{\overline m}$ for the latitude of Casablanca, Morocco ($33^\circ 32'$ N).\la{fig:zenith}}
\end{figure}

\subsection{Correcting for size of solar disk and atmospheric refraction}
\la{sec:adjustments}

Comparison of the results of \Eq{eq:hours} with the hours of sunrise and sunset published in newspapers and other sources reveals a discrepancy: we underestimate the duration of daylight by several minutes.  One reason for this discrepancy is that the times of sunrise and sunset are usually taken to correspond to the moments when the upper edge of the solar disk crosses the horizon.  Since the solar disk has an angular diameter of about $0.5^\circ$, sunrise occurs slightly before, and sunset slightly after, the times that we have computed, which referred to the horizon crossings of the Sun's center.\footnote{The reader might care to estimate the percent change in the angular diameter of the solar disk caused by the yearly variation in the distance between the Earth and the Sun, keeping in mind that the eccentricity of an elliptic orbit may be expressed as \hbox{$e = (r_a - r_p)/(r_a + r_p)$, where $r_{a,p}$} are the orbital radii at aphelion and perihelion, respectively (see \Fig{fig:orbit}), and that $e = 0.0167$ for the Earth.}

An even more important factor is that, since the density of the Earth's atmosphere is greater closer to the Earth's surface, light passing obliquely through the atmosphere bends downwards.  This atmospheric refraction {\it lifts} the Sun's apparent position when it is near the horizon.\footnote{The physics responsible for atmospheric refraction at low altitudes ($\alpha \sim 0$) has recently been analyzed in \cite{refraction-physics}.}  When the Sun (or any other celestial object) appears to us to be at horizon, its true altitude is about $-0.6^\circ$ \cite{Meeus-rising}.  For a general treatment of atmospheric refraction and other effects on the apparent position of celestial objects as seen from the surface of the Earth, see \cite{geocentric}.

In \cite{Saemundsson}, Saemundsson proposed a simple formula for the apparent shift in altitude, $\Delta \alpha$, as a function of the true altitude $\alpha$, accurate to within $0.07'$ for all values of $\alpha$:
\be
\Delta \alpha = \frac{1.02'}{\tan \left( \alpha + \frac{10.3^\circ}{\alpha + 5.11^\circ} \times 1^\circ \right)}~.
\la{eq:refraction}
\ee
This expression may be used to adjust the altitude of \Eq{eq:altitude} in order to bring it into closer agreement with observation; see also \cite{refraction-formulas}.

Combining the size of the solar disk with the lifting due to refraction, we see that sunrise and sunset correspond to the moments when the true altitude of the center of the solar disk is \hbox{$\alpha_\odot = - \left(0.5^\circ / 2 + 0.6^\circ \right) = - 0.85^\circ$}.  Thus, it is necessary to adjust the diagrams of \Fig{fig:daylight} to account for this displacement, in order to bring our expression for the duration of daylight into agreement with the interval between the published times of sunrise and sunset.

\begin{figure} [t]
\begin{center}
	\subfigure[]{\includegraphics[width=0.355 \textwidth]{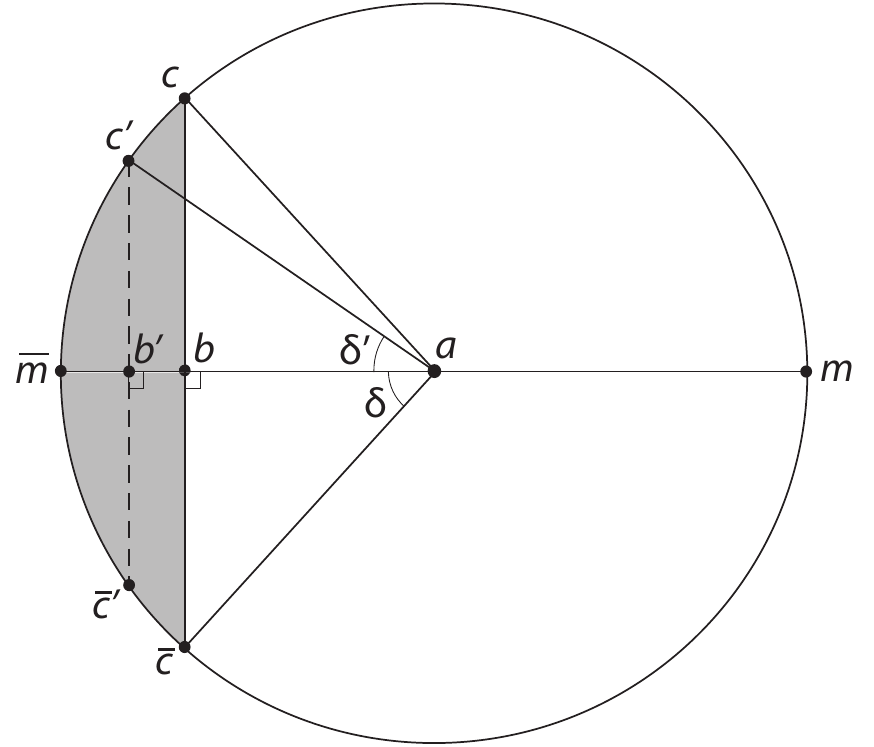}} \hskip 1.5 cm
	\subfigure[]{\includegraphics[width=0.5 \textwidth]{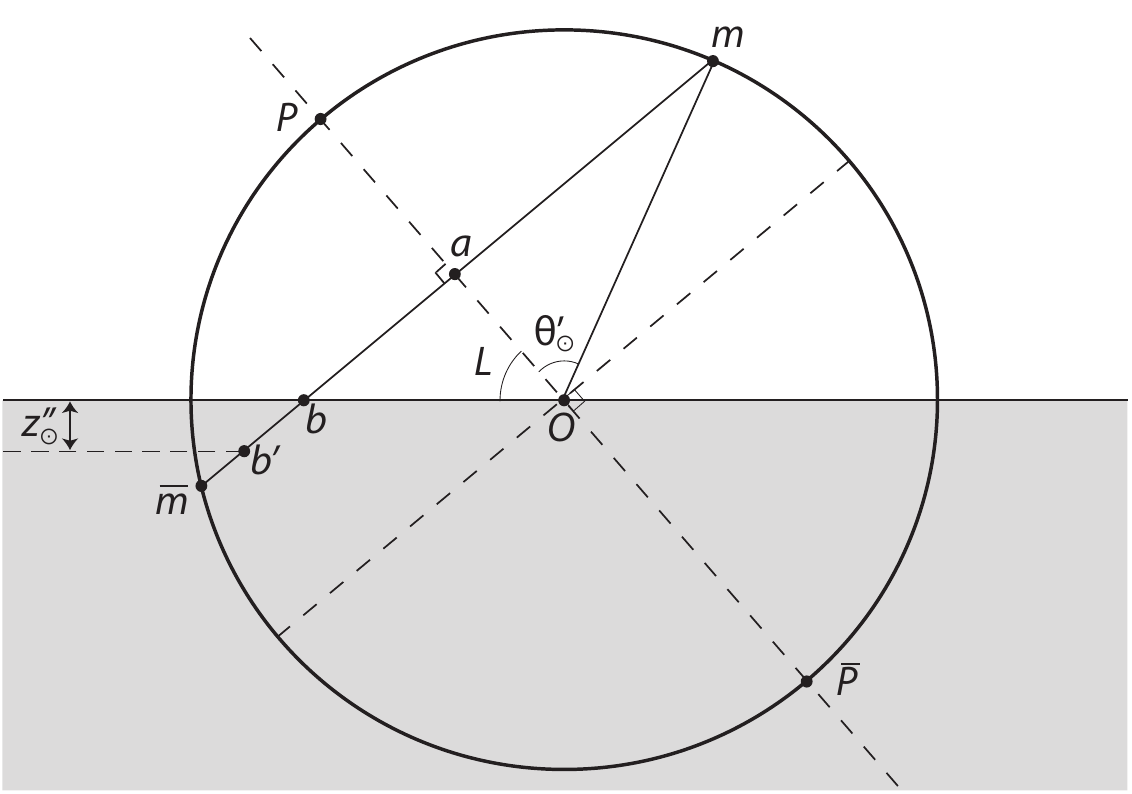}}
\end{center}
\caption{\small (a): Cross-section of the celestial sphere, centered at point $a$ and perpendicular to the Earth's axis of rotation $P \overline P$; see \Fig{fig:daylight}(b).  The points $c'$ and $\bar c'$ correspond to sunrise and sunset, respectively, after taking into account the size of the solar disk and the effect of atmospheric refraction.  (b): Cross-section of the celestial sphere, along the Earth's axis of rotation and centered at the observer's position $O$; see \Fig{fig:daylight}(a).  The vertical displacement between $b$ and $b'$ corresponds to the terrestrial coordinate $z''_\odot$ (see \Eq{eq:terrestrial}) at the time of sunrise or sunset.\la{fig:adjustment}}
\end{figure}

Points $c'$ and $\bar c'$ in \Fig{fig:adjustment}(a) mark the position of the Sun at the published times of sunrise and sunset, respectively.  Point $b'$ marks the projection of these unto the line $m \overline m$.  The correction to the duration of daylight, in hours, is therefore
\be
\Delta H = 24 \left( \frac{\delta - \delta'}{\pi} \right)~.
\la{eq:adjustment-delta}
\ee
In order to obtain an expression for the adjustment $\Delta H$ we therefore need to express $\delta - \delta'$ as a function of the latitude and the date of the year.

The vertical direction in \Fig{fig:adjustment}(b) corresponds to $z''$-axis in the terrestrial reference frame of \Eq{eq:terrestrial}.  The vertical displacement between $b$ and $b'$ therefore corresponds to the value of $z''_\odot$ at the published times of sunrise or sunset, so that
\be
bb' = \frac{\left| z''_\odot \right|}{\cos L} = \frac{\sin 0.85^\circ}{\cos L} = \frac{0.015}{\cos L}~.
\la{eq:adjustment-vertical}
\ee
Meanwhile, we can see from the diagrams in \Fig{fig:adjustment} that 
\be
bb' = ab' - ab = \sin \theta'_\odot \left( \cos \delta' - \cos \delta \right)
\simeq \sin \theta'_\odot \sin \delta \left( \delta - \delta' \right)~.
\la{eq:adjustment-oblique}
\ee

Combining Eqs.~(\ref{eq:adjustment-delta}), (\ref{eq:adjustment-vertical}), and (\ref{eq:adjustment-oblique}), we obtain
\be
\Delta H = \frac{7 ~\hbox{min.}}{\cos L \sin \theta'_\odot \sin \delta}~.
\la{eq:adjustment-final}
\ee
We can express the value of $\Delta H$ as a function of latitude $L$ and date $d$ by plugging in the expression for $\theta'_\odot$ in \Eq{eq:polar-sun} and for $\delta$ in \Eq{eq:delta}.  Equation (\ref{eq:adjustment-final}) implies that the adjustment to the duration of daylight from the size of the solar disk and atmospheric refraction amounts to at least 7 minutes, for all locations and dates of the year, and can be significantly larger for locations distant from the Equator ($L = 0$) and on dates far from the equinoxes (when $\theta'_\odot = \delta = \pi / 2$).\footnote{Equation (\ref{eq:adjustment-final}) should not be used at Arctic or Antarctic latitudes (i.e., $| L | \geq \pi / 2 - \varepsilon$), or very near them, because $(\delta - \delta')/\delta$ might not always be small, invalidating the approximation in \Eq{eq:adjustment-oblique}, and also because $\alpha_m$ and $\alpha_{\overline m}$ may have the same sign, implying that the Sun neither rises nor sets.}

Note also that the position of the Sun with respect to the horizon is further lifted, and the duration of daylight consequently lengthened, by observing from an elevated position, since this makes the viewer's horizon recede.  Indeed, it is possible to estimate the size of the Earth from the time difference between the sunset over the ocean as seen by an observer lying on the ground and the sunset as seen by the same observer standing up; see \cite{HR-sunset}.

\section{Solar alignments}
\la{sec:alignments}

Stonehenge, in the English county of Wiltshire, was built in prehistoric times in such a way that around the time of the summer solstice an observer standing at the center of the circular structure sees the Sun rise above the outer Heelstone.  There are other famous instances of ancient structures oriented according to solar alignments.  The Great Temple of Amen-Ra, at Karnak, Egypt, was designed so that the last rays of the Sun on the day of the summer solstice illuminate the inner sanctuary.  The Mayan pyramid of Kukulk\'an (also called {\it El Castillo}, ``The Castle'') in Chich\'en Itz\'a, in the Mexican state of Yucat\'an, was built so that the setting Sun, around the time of the equinoxes, casts a shadow that looks like a serpent slithering down the side of the staircase.  Modernly, the date and vantage point of Monet's series of paintings of the British Houses of Parliament have been determined from the position of the Sun in the pictures \cite{Monet}.  Such alignments can be readily studied by the methods developed in this article.

\subsection{Direction to sunrise and sunset}
\la{sec:sunrise-position}

\begin{figure} [t]
\begin{center}
	\includegraphics[width=0.35 \textwidth]{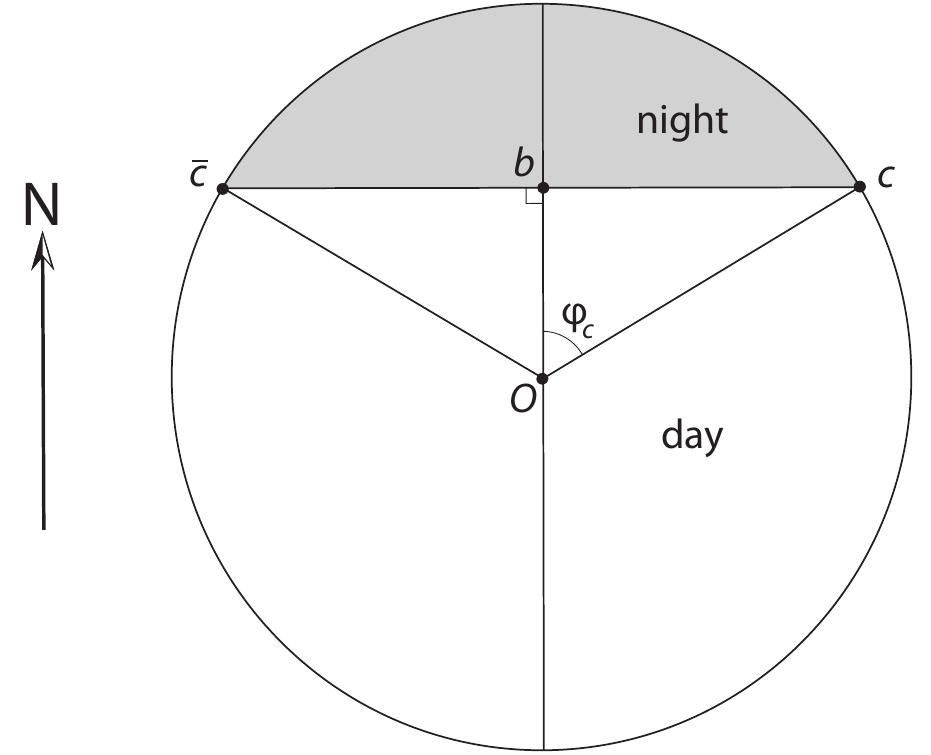}
\end{center}
\caption{\small Cross-section of the celestial sphere ---as shown in \Fig{fig:daylight}(a)--- along the horizontal plane passing through the position of the observer at $O$.  The ray from $O$ to $b$ points North for an observer in the northern hemisphere (South for an observer in the southern hemisphere).  The Sun rises at $c$ and sets at $\bar c$.  The angle $\varphi_c$ gives the geographic azimuth of sunrise.\la{fig:sunrise-position}}
\end{figure}

Knowing $\theta'_\odot$, we can easily compute the direction, with respect to the cardinal points, in which an observer will see the Sun rise and set.  Figure \ref{fig:sunrise-position} shows a cross-section of the celestial sphere, this time through the horizontal plane passing through the position of the observer at $O$, so that the resulting circle has the same unit radius as the celestial sphere itself.  If the observer is on the northern hemisphere, then the ray from $O$ to $b$ points North (whereas it points South for an observer in the southern hemisphere).

The angle $\varphi_c$ therefore gives the geographic azimuth (i.e., the compass bearing) of sunrise and can be expressed as:
\be
\varphi_c = \arccos{Ob} = \arccos \left( \frac{\cos \theta'_\odot}{\cos L} \right)~.
\la{eq:sunrise-azimuth}
\ee
(This agrees with the result given in \cite{Khavrus}.)  If we neglect the change in the position of the Sun between sunrise and sunset on the same day, then the azimuth of the sunset at $\bar c$ is simply $\varphi_{\bar c} = 2 \pi - \varphi_c$.

\subsection{Manhattanhenge}
\la{sec:Manhattan}

\begin{figure} [t]
\begin{center}
	\includegraphics[width=0.42 \textwidth]{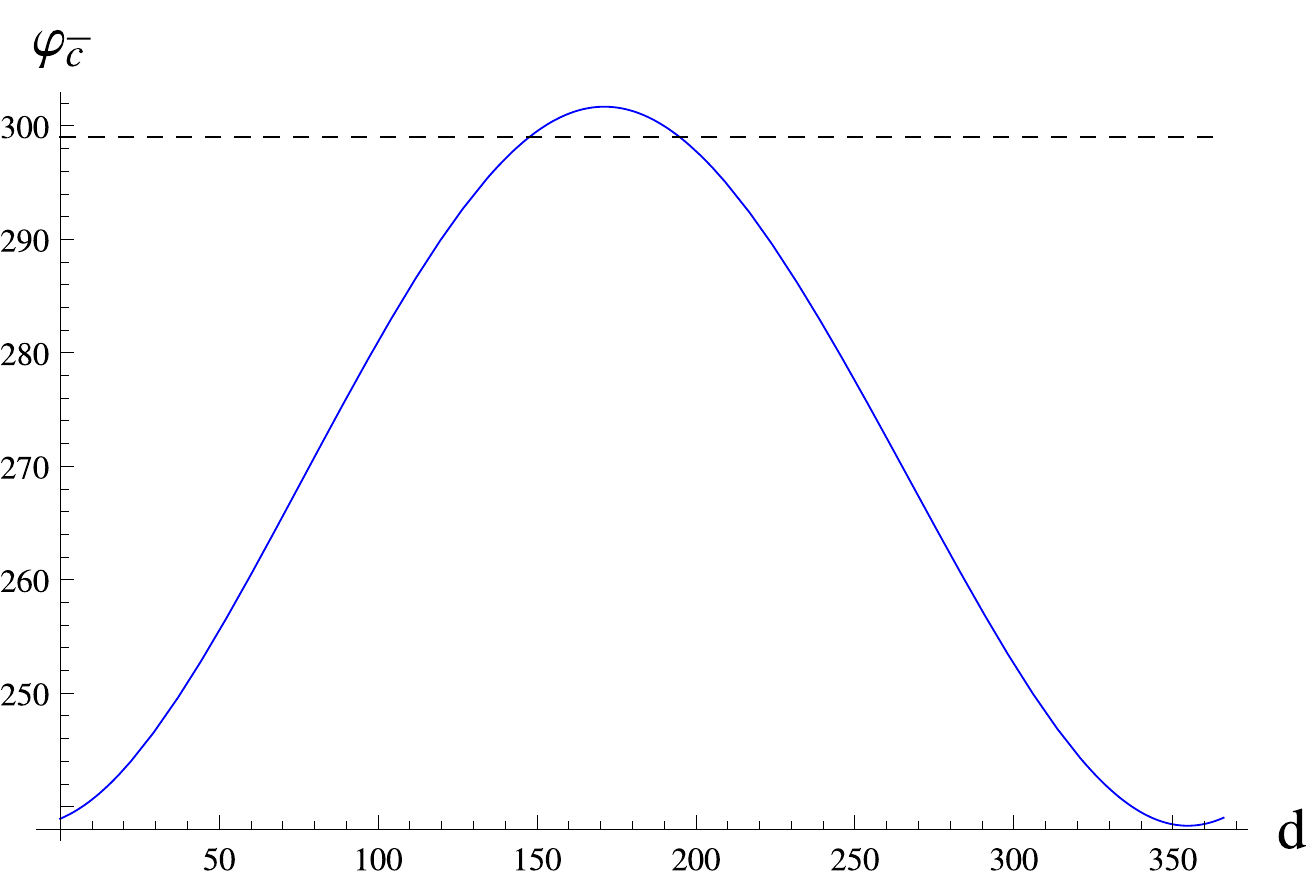}
\end{center}
\caption{\small Plot of the geographic azimuth for sunset, $\varphi_{\bar c}$, in degrees, at the latitude of Manhattan, in New York City ($40^\circ 47'$ N), as a function of the day of the year $d$ (starting on 1 January).  When $\varphi_{\bar c} = 299^\circ$ (marked by the dashed line) the sunset is aligned with the east-west streets on the main traffic grid.\la{fig:manhattan-sunset}}
\end{figure}

Figure \ref{fig:manhattan-sunset} gives a plot of the sunset's azimuth $\varphi_{\bar c}$, as a function of the date $d$, for the latitude of Manhattan.  This can be used to find the dates of ``Manhattanhenge'' (also called the ``Manhattan solstice''), when a New York City pedestrian can see the sunset in between the skyscrapers, because the sunset is aligned with the east-west streets on the main traffic grid for the borough of Manhattan \cite{Manhattanhenge}.  Since those streets point $29^\circ$ north of true west, this alignment occurs when $\varphi_{\bar c}(d) = 270^\circ + 29^\circ = 299^\circ$.  For the year 2013 these correspond to 28 May ($d = 147$) and 14 July ($d=194$).

A more impressive visual spectacle than the alignment of the actual sunset with the east-west streets is the observation of the full solar disk in between the profiles of the buildings and slightly above the horizon.  This occurs a couple of days {\it after} the first date of the year for which $\varphi_{\bar c}(d) = 299^\circ$, as well as a couple of days {\it before} the second such date (see \cite{Manhattanhenge}).  Figure \ref{fig:EmpireState} shows a photograph of the view on 11 July 2012, looking west along 34th Street in midtown Manhattan, several minutes before sunset.

\begin{figure} [t]
\begin{center}
	\includegraphics[width=0.4 \textwidth]{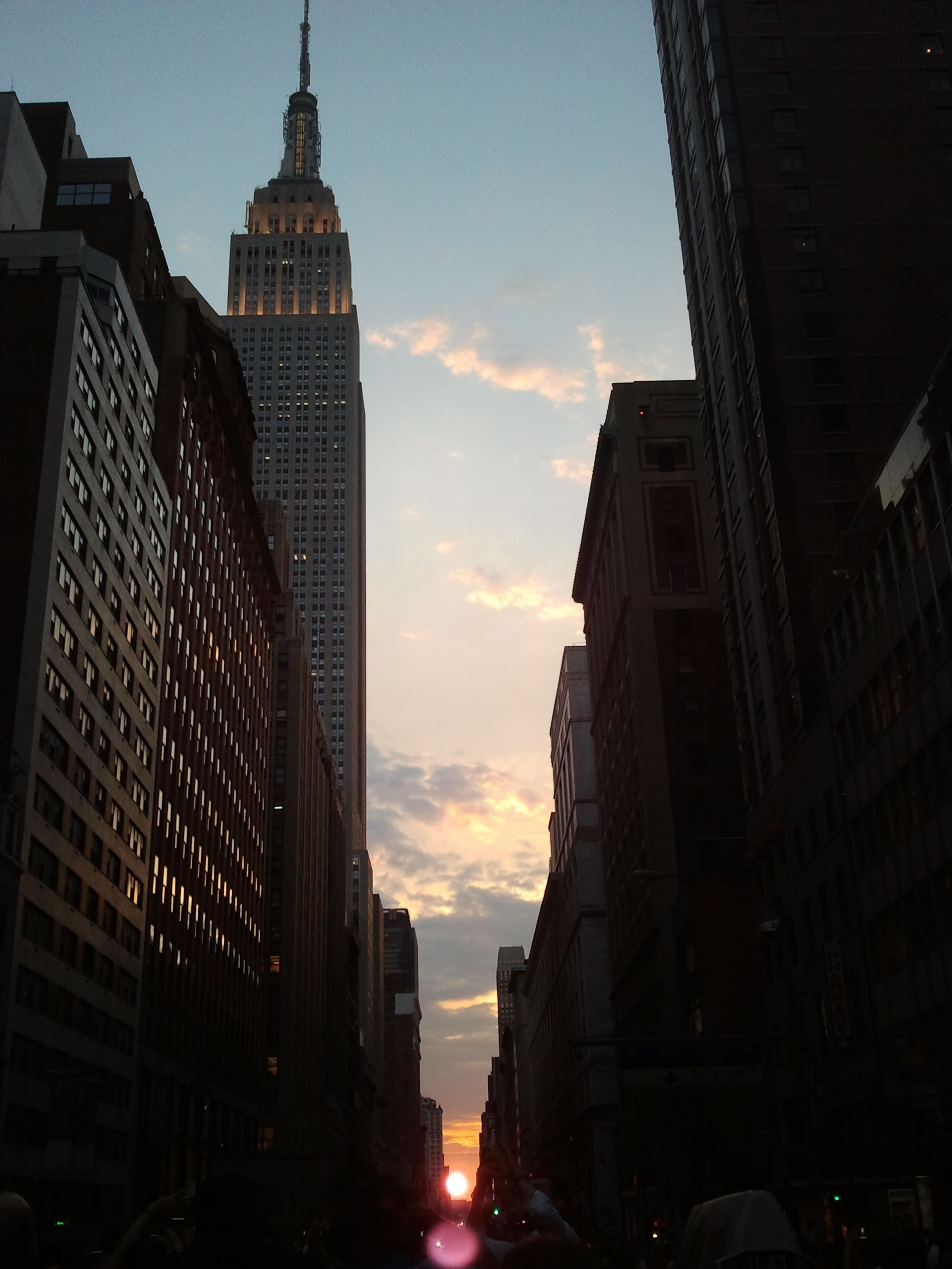}
\end{center}
\caption{\small Photograph taken by the author on 11 July 2012, at 8:17 p.m., Eastern Standard Time (EST), looking west along 34th Street, from a position near the intersection with Park Avenue, in midtown Manhattan, New York City.  The solar disk can be seen between the profiles of the buildings, a few degrees above the horizon.  The Empire State Building is prominent on the left.\la{fig:EmpireState}}
\end{figure}

\section{Geographic longitude}
\la{sec:longitude}

There is one issue left to resolve before we can give the Sun's position in the sky for a given time and location: the choice of $t_0$ in \Eq{eq:altitude}, needed to translate from the $\phi''_\odot$ in \Eq{eq:terrestrial} to a geographic azimuth $\varphi_\odot$ defined with respect to the cardinal points (North, East, South, and West).  This must be consistent with how $t$ is measured in \Eq{eq:mean-numbers}.  The choice of $t_0$ will evidently depend on the observer's geographic longitude, since at a given universal time the azimuth of the Sun with respect to the cardinal points depends on the longitude at that location.

\subsection{Reference time}
\la{sec:reference}

At the time of the northward equinox, when $\phi_\odot = 0$, the Sun's altitude, observed at the Equator ($L=0$), is given by \Eq{eq:altitude} as $\alpha_\odot = \omega \cdot (t_0 - t)$.  Therefore, we can simply find the geographic longitude $\ell_0$ at which the equinox occurs {\it precisely at sunset} and, for any other longitude $\ell$, choose:
\be
t_0 (\ell)= t_{\rm eq} - \frac{\ell - \ell_0}{\omega} ~,
\la{eq:longitude}
\ee     
where $t_{\rm eq}$ is the time of the northward equinox and $\omega$ is the rate of the Earth's rotation (see \Eq{eq:omega}).  The longitude $\ell$ is taken to be positive to the {\it east} of the Prime Meridian.

For instance, the first equinox for the year 2013 occurs on 20 March at 11:02 UTC \cite{equinox-time}, so that $t_{\rm eq} = 78.46$ days.  At that moment, the difference between apparent solar time and the ``mean solar time'' (which matches UTC for $\ell = 0$) is about 7 minutes \cite{eq-time}.  Thus, the mean solar time for sunset at the time of the equinox is 18:07.\footnote{The variation in the difference between apparent solar time and mean solar time throughout the year is given by the ``equation of time.''  We shall return to this issue in \Sec{sec:analemma}.}  The first equinox of 2013 therefore occurs at sunset at the geographic longitude of
\be
\ell_0 = \frac{\hbox{18:07} - \hbox{11:02}}{\hbox{24:00}} \times 2 \pi = 1.854 = 106.3^\circ ~,
\ee
to the east of the Prime Meridian.

\subsection{Solar azimuth}
\la{sec:azimuth}

Note that $\phi''_\odot$ {\it decreases} during the course of a day, and that we have chosen $t_0$ in \Eq{eq:terrestrial} so that $\phi''_\odot (L,\ell_0,t_{\rm eq}) = 0$.  Sunset at the time of the equinox must point directly West.  Conventionally, the geographic azimuth $\varphi$ is defined so that $\varphi = 0$ corresponds to North, $\varphi = 90^\circ$ to East, $\varphi = 180^\circ$ to South, and $\varphi = 270^\circ$ to West.  Therefore we take
\be
\varphi_\odot = 270^\circ - \phi''_\odot ~.
\ee

\section{Altitudes}
\la{sec:altitudes}

We shall now plot the altitude of the Sun above the horizon as a function of time (measured over the course of one year), for several geographic locations of interest.  These are the results obtained from Eqs.~(\ref{eq:altitude}), (\ref{eq:center-numbers}), (\ref{eq:mean-numbers}), and (\ref{eq:longitude}), for the center of the solar disk.\footnote{These plots were made without the correction for atmospheric refraction of \Eq{eq:refraction}, which in any case is far too small an adjustment to be discernible at the resolution of our graphs.}

\subsection{Buenos Aires, Argentina}
\la{sec:buenosaires}

Buenos Aires, the capital of Argentina, is located a latitude $34^\circ 36'$ South, longitude $58^\circ 23'$ West.  The pattern of the Sun's altitude here, shown in \Fig{fig:buenosaires}, is typical of southern temperate regions, with longer days around the beginning and end of the calendar year, and shorter days around midyear.

\begin{figure} [h]
\begin{center}
	\includegraphics[width=\textwidth]{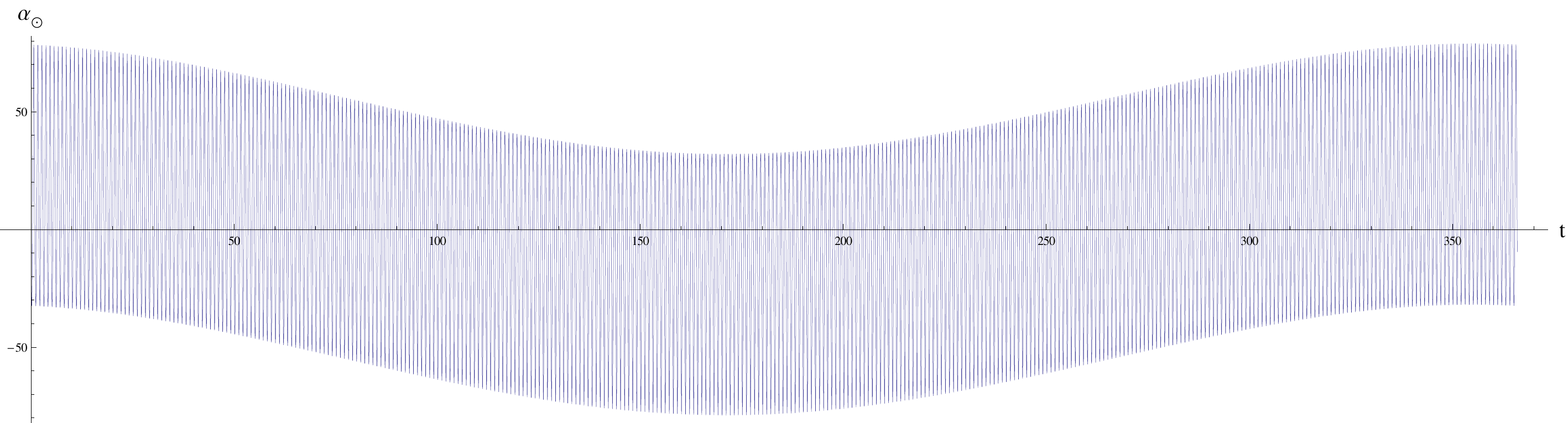}
\end{center}
\caption{\small Altitude of the Sun above the horizon, $\alpha_\odot$, in degrees, at the coordinates of Buenos Aires, Argentina ($34^\circ 36'$ S, $58^\circ 23'$ W), as a function of the time of the year $t$ (measured in days from 1 January, 2013, 0:00 UTC).\la{fig:buenosaires}}
\end{figure}

\subsection{Alert, Nunavut, Canada}
\la{sec:alert}

Alert, in the Canadian territory of Nunavut, is the northernmost permanently inhabited place on Earth, located 817 km from the geographic north pole.  Its geographic coordinates are latitude $82^\circ 30'$ North, longitude $62^\circ 20'$ West.  Here the Sun rises and sets only in March and April, and again in September and October.  During the rest of the year the Sun remains either above or below the horizon, as shown in \Fig{fig:alert}.

\begin{figure} [h]
\begin{center}
	\includegraphics[width=\textwidth]{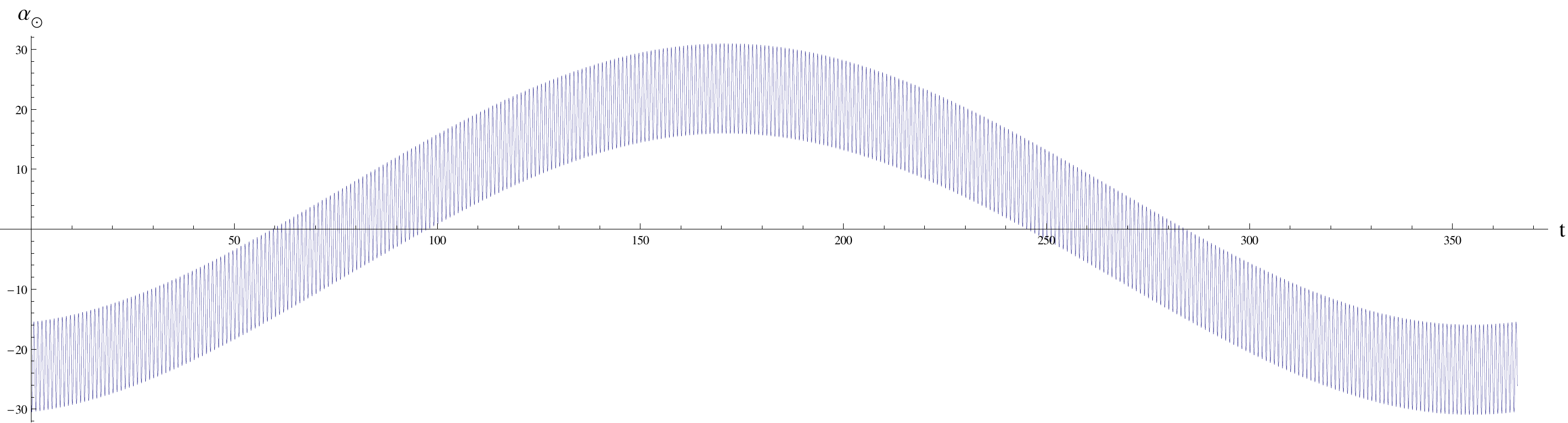}
\end{center}
\caption{\small Altitude of the Sun above the horizon, $\alpha_\odot$, in degrees, at the coordinates of Alert, Nunavut, Canada ($82^\circ 30'$ N, $62^\circ 20'$ W), as a function of the time of the year $t$ (measured in days from 1 January, 2013, 0:00 UTC).\la{fig:alert}}
\end{figure}

\subsection{Singapore}
\la{sec:singapore}

Singapore, a city coextensive with the independent nation of the same name, is located very near the Equator, at latitude $1^\circ 17'$ North, longitude $103^\circ 50'$ East.  Here the days remain almost evenly divided between light and darkness.  The Sun only reaches zenith at around the times of the equinoxes (in late June and late September), as shown in \Fig{fig:singapore}.

\begin{figure} [h]
\begin{center}
	\includegraphics[width=\textwidth]{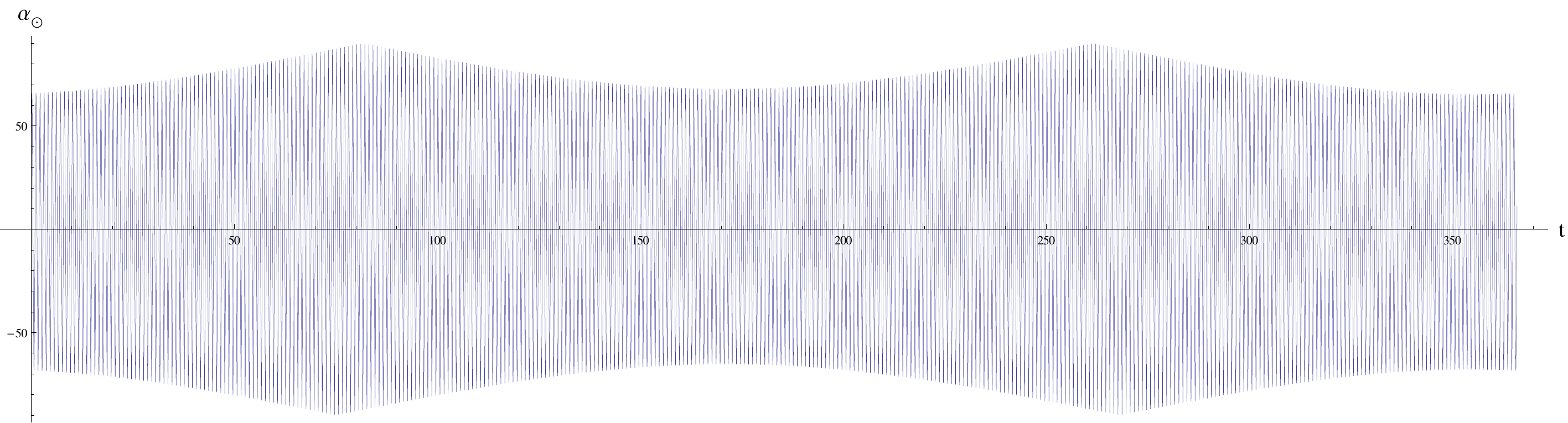}
\end{center}
\caption{\small Altitude of the Sun above the horizon, $\alpha_\odot$, in degrees, at the coordinates of Singapore ($1^\circ 17'$ N, $103^\circ 50'$ E), as a function of the time of the year $t$ (measured in days from 1 January, 2013, 0:00 UTC).\la{fig:singapore}}
\end{figure}

\subsection{San Jos\'e, Costa Rica}
\la{sec:sanjose}

San Jos\'e, the capital of Costa Rica, is located at latitude $9^\circ 56'$ North, longitude $84^\circ 5'$ West.  The pattern of the Sun's altitude, shown in \Fig{fig:sanjose}, is typical of northern tropical regions, with slightly longer days around the midyear solstice.  Note that the noonday Sun reaches the zenith around two different dates, which by \Eq{eq:a-maxmin} are given by
\be
\theta'_\odot (d) = 90^\circ - 9^\circ 56' = 80.07^\circ = 1.397~.
\ee
Using \Eq{eq:polar-sun}, we find that the solutions for the year 2013 are $d = 104$ (corresponding to 15 April) and $d = 238$ (corresponding to 27 August).

\begin{figure} [h]
\begin{center}
	\includegraphics[width=\textwidth]{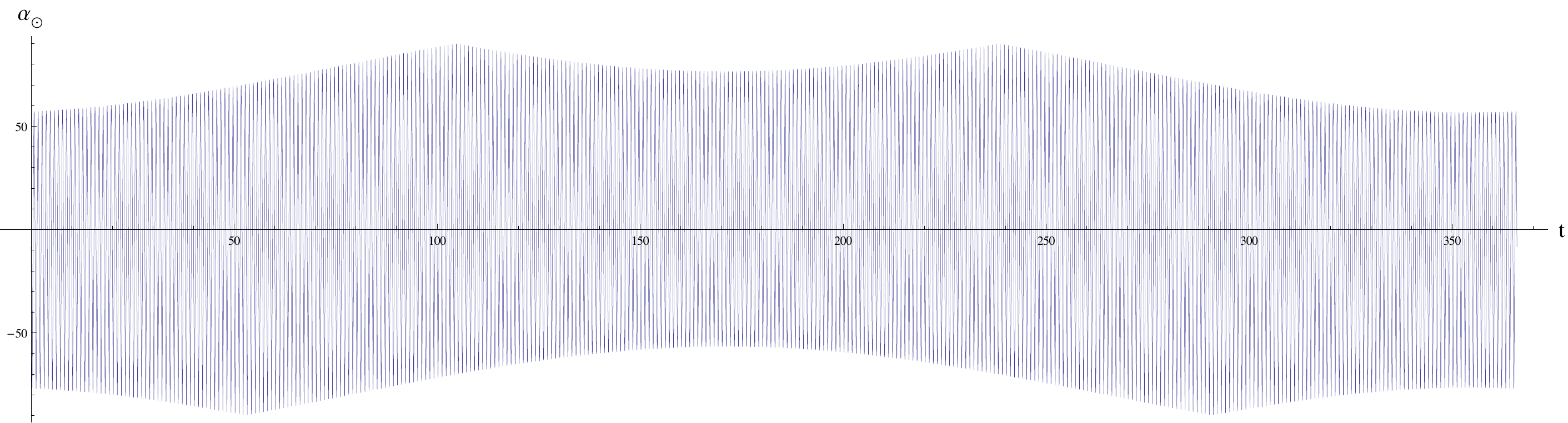}
\end{center}
\caption{\small Altitude of the Sun above the horizon, $\alpha_\odot$, in degrees, at the coordinates of San Jos\'e, Costa Rica ($9^\circ 56'$  N, $84^\circ 5'$ W), as a function of the time of the year $t$ (measured in days from 1 January, 2013, 0:00 UTC).\la{fig:sanjose}}
\end{figure}
¶
\section{Analemmas and equation of time}
\la{sec:analemma}

The mean solar time is defined in terms of a fictitious ``mean sun,'' which moves at a uniform rate along the celestial equator.  This differs from the actual position of the Sun, which moves along the ecliptic at a rate not quite uniform (see, e.g., \cite{meansun}).  At noon, the fictitious mean sun crosses the {\it meridian} above the observer. Here, the word ``meridian'' is understood not as a line of constant geographic longitude, but rather as the great circle $OP$ in \Fig{fig:daylight}(a), which is perpendicular to the observer's horizon and passes through the celestial poles.  The hour of 12:00 UTC corresponds to the time when the mean sun, as seen by an observer at any location along the Prime Meridian (longitude $\ell = 0$), has a positive altitude and a geographic azimuth of either $\varphi = 180^\circ$ (for observers in the northern hemisphere) or $\varphi = 0$ (for observers in southern hemisphere).

If we plot the {\it actual} Sun's altitude $\alpha_\odot$ vs. its azimuth $\varphi_\odot$, over the course of many days, for a given mean solar time and at a given location, the result will look like a figure-eight, called an {\it analemma}.  Figure \ref{fig:analemma} illustrates this for the geographic coordinates of Greenwich, England ($51^\circ 29'$ N, $0^\circ$ E).  These plots correspond, respectively, to fixing the hour of observation at 0:00, 6:00, 12:00, and 18:00 UTC, for each day of the year 2013.

The difference, as a function of the date of the year, between noon and the actual time when the Sun crosses the meridian, is called the ``equation of time.''  This difference results mainly from the projection of points along the ecliptic onto the equatorial plane (i.e., from the coordinate transformation between the ecliptic and equatorial frames, given by \Eq{eq:equatorial}, which implies that $\tan \phi'_\odot = \cos \varepsilon \tan \phi_\odot$, distorting $\phi'_\odot$ with respect to $\phi_\odot$), and to a lesser extent also from the eccentricity of the Earth's orbit, which makes the motion of the Sun along the ecliptic, described by $\phi_\odot$, less than perfectly uniform (see \Sec{sec:center}).\footnote{Introductory textbooks and other pedagogical sources sometimes explain the analemma's horizontal displacement as resulting solely or primarily from the eccentricity of the Earth's orbit (see, e.g., \cite{Dixon-analemma}).  This is clearly incorrect:  Kepler's second law implies that a planet's orbital angular velocity is faster than the mean when the planet is near the perihelion and slower when it is near the aphelion (see \Fig{fig:orbit}).  If this were the dominant factor in determining the equation of time, then the azimuth of the Sun, observed at noon, would be greater than $180^\circ$ during half of the year and smaller than $180^\circ$ during the other half, giving a figure-zero rather than a figure-eight for the analemma.  Orbital eccentricity does play an important role in determining the precise form of the equation of time, but the analemma would still be a figure-eight if the Earth's orbit were perfectly circular.}

The equation of time can be read from the horizontal displacement of the analemma in \Fig{fig:analemma}(c).  In that plot, each degree of azimuthal displacement away from $\varphi_\odot = 180^\circ$ translates to 24 hours $/360 = 4$ minutes in the equation of time for that date.  A convenient check of the validity of our computations is to compare this analemma with the one available at \cite{Wikipedia-analemma}, based on data from \cite{JPL}.  

\begin{figure} [t]
\begin{center}
	\subfigure[]{\includegraphics[width=0.45 \textwidth]{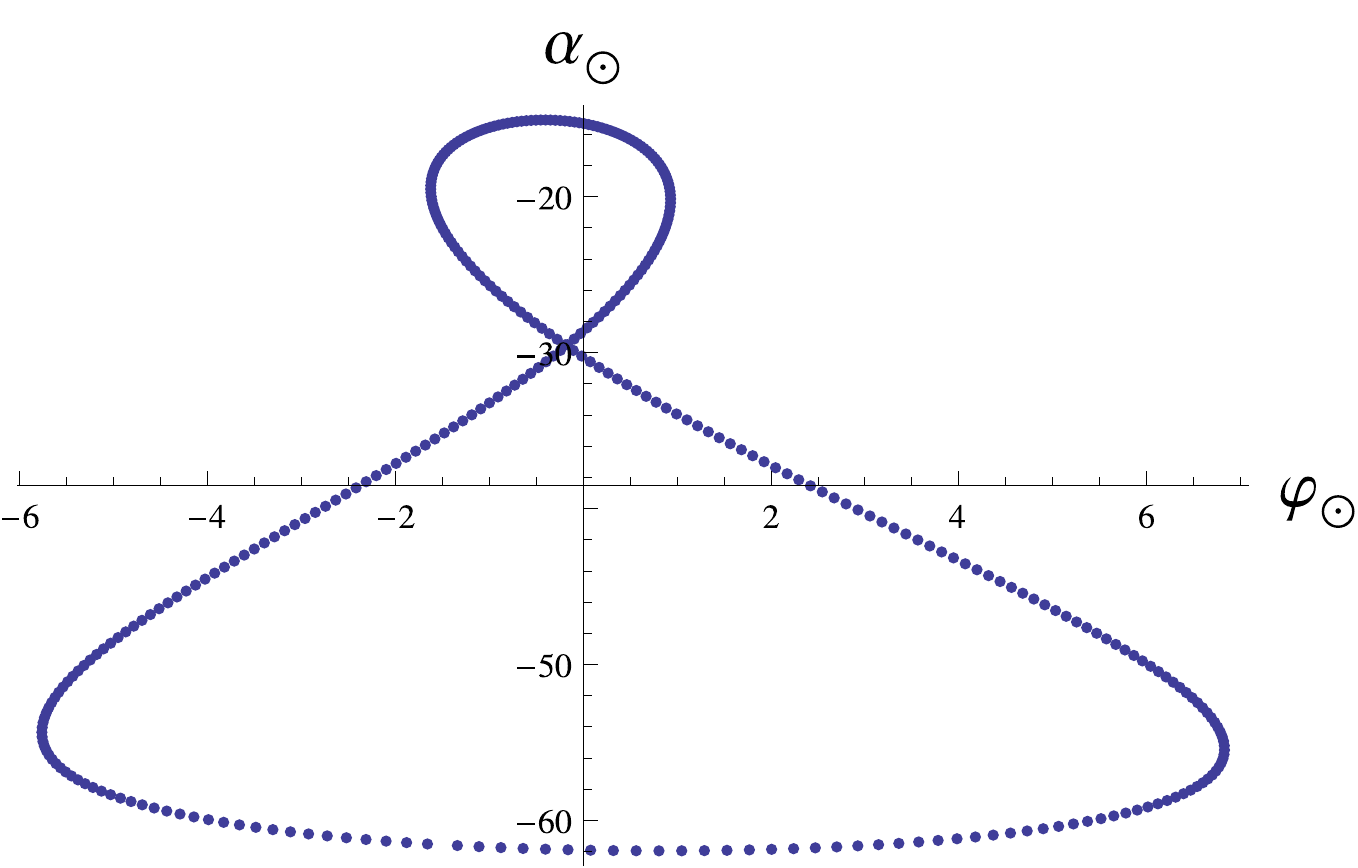}} \hskip 1 cm
	\subfigure[]{\includegraphics[width=0.45 \textwidth]{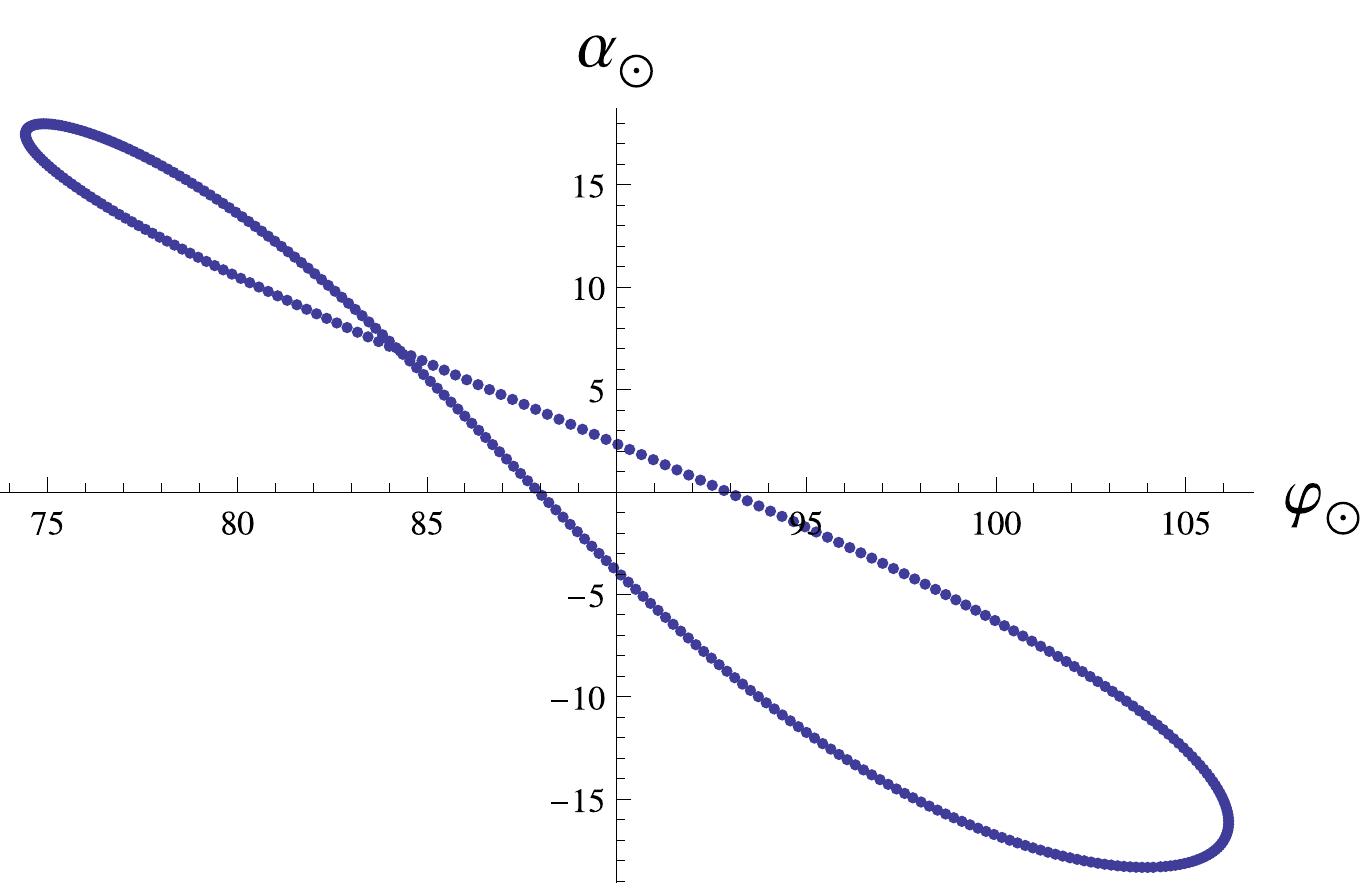}} \\
	\subfigure[]{\includegraphics[width=0.45 \textwidth]{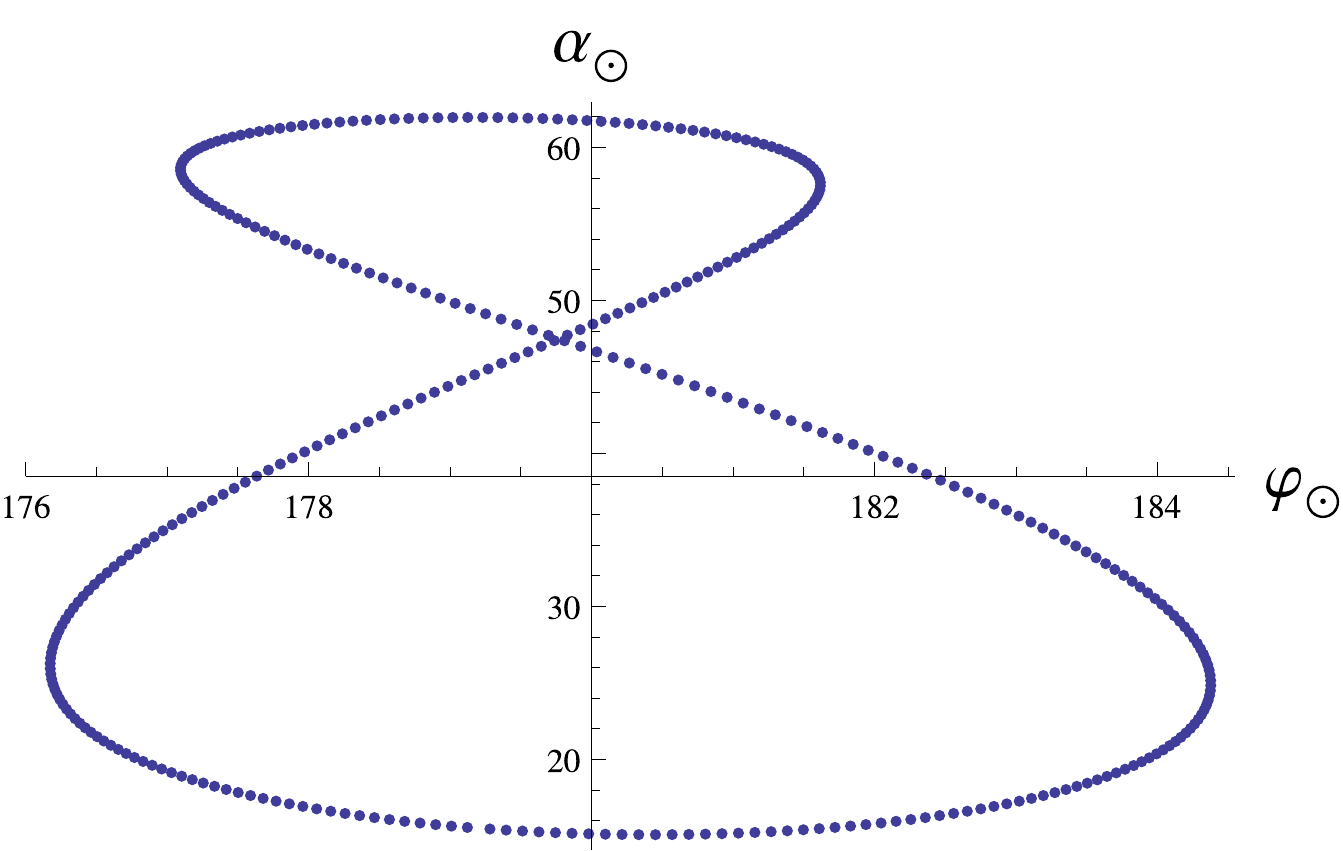}} \hskip 1 cm
	\subfigure[]{\includegraphics[width=0.45 \textwidth]{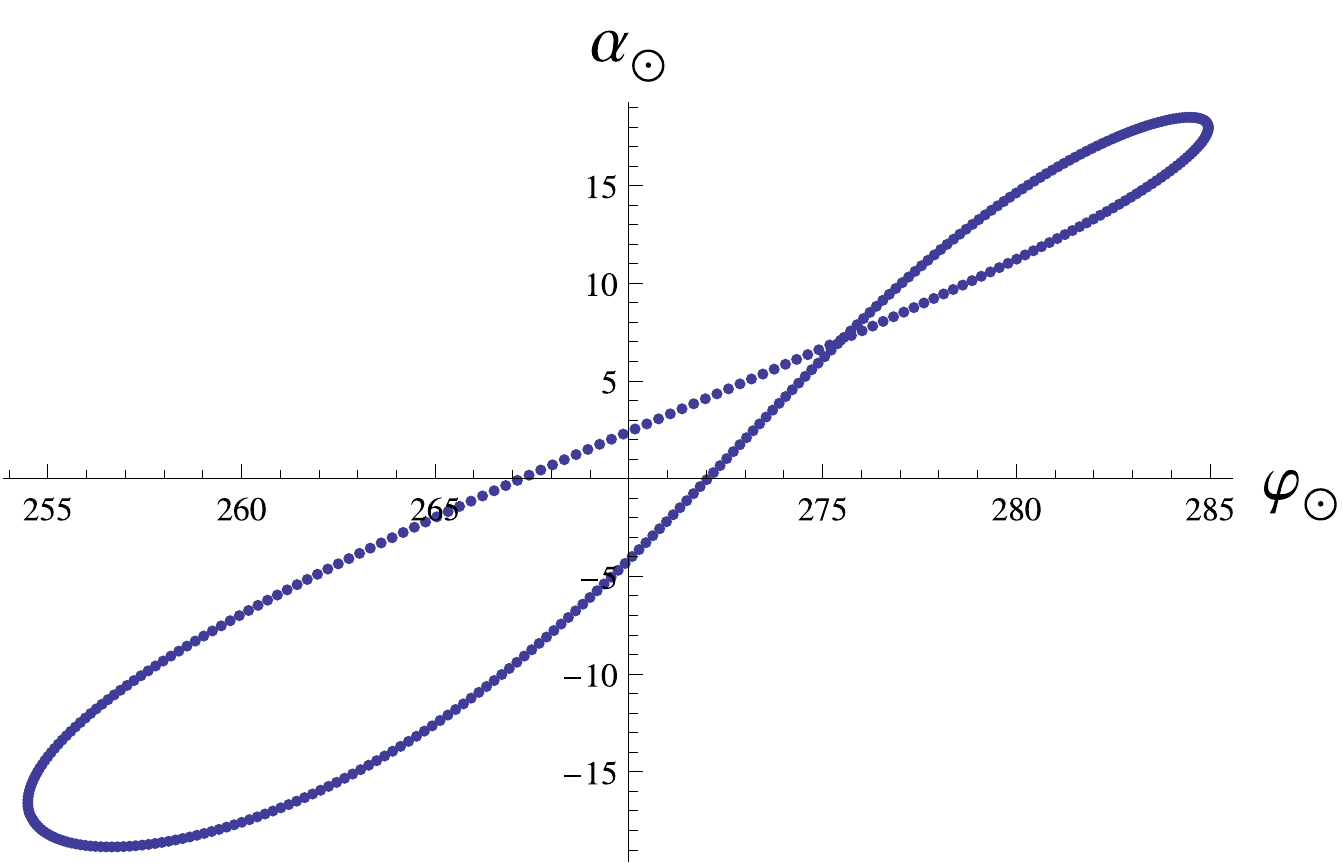}}
\end{center}
\caption{\small Plots of solar altitude $\alpha_\odot$ vs. azimuth $\varphi_\odot$, in degrees, for each day of 2013, at the geographic coordinates of Greenwich, England ($51^\circ 29'$ N, $0^\circ$ E), for the hours of: (a) 0:00, (b) 6:00, (c) 12:00, and (d) 18:00 UTC.\la{fig:analemma}}
\end{figure}

Even though the position of the Sun in the sky for a given hour, calendar date, and location, is not quite constant from one year to the next, the way in which the mean solar time is defined ensures that the shape of the analemma remains unchanged.  The word ``analemma'' derives from the Greek term for the pedestal of a sundial and refers to the instructions on how to correct a sundial's reading for the equation of time, which can give an adjustment of as much as a quarter of an hour.  For a practical treatment of these issues as they relate to the construction and reading of sundials, see \cite{Waugh}.  The equation of time has been investigated analytically in \cite{Muller,Vandyck,Goyder}.

\section{Discussion}
\la{sec:discussion}

The derivation in \Sec{sec:spherical}, based on performing three successive coordinate rotations, should be easier to follow, without introducing approximations, than the analytic treatments of solar position currently available in the literature.  The material that we have presented here thus offers opportunities for discussion and investigation by students with only prior knowledge of planar geometry and elementary linear algebra.  Our approach also gives us the freedom to incorporate the correction from the eccentricity of the Earth's orbit (see \Sec{sec:astronomical}) or not, depending on the level of accuracy and sophistication sought.  The derivations of \Sec{sec:daylight} and \Sec{sec:alignments}, expressed in terms of the angle $\theta'_\odot$, require only planar geometry and can be used independently of the linear algebra of \Sec{sec:spherical} (which is the approach taken in \cite{Khavrus}).

To the beginner in astronomy, this work may also serve to introduce or reinforce the concept of the celestial sphere, and of the ecliptic and equatorial coordinates therein.  In \Sec{sec:analemma}, we had occasion to correct a misconception about the equation of time that has often been repeated in the pedagogical literature.  The results in this article may also be used to compute variables relevant to solar-powered technology, as stressed in \cite{Probst,Sproul,Khavrus}.  The {\it Mathematica} notebook that accompanies this article can be a resource for relatively advanced students interested in the subject.

\begin{acknowledgements}

I thank my teachers Charles McElroy and Jos\'e Alberto Villalobos for interesting me in this problem, many years ago.  Mr.~Villalobos also kindly provided feedback on this manuscript and pointed me to \cite{Meeus}.  I thank Take Okui for calling my attention to the effect of atmospheric refraction on the duration of daylight, and Felice Kuan and Graeme Smith for hospitality during the expedition to observe the second Manhattanhenge of 2012.  I thank an anonymous referee for pointing out to me \cite{Khavrus} and other pedagogical articles related to this subject, and for suggestions on improving this manuscript.

\end{acknowledgements}


\bibliographystyle{aipprocl}   

\end{document}